\documentclass[11pt]{article}
\usepackage{aaspp4}
\tighten

\def\gtsima{$\, \buildrel > \over \sim \,$}
\def\ltsima{$\, \buildrel < \over \sim \,$}
\def\simgt{\lower.5ex\hbox{\gtsima}}
\def\simlt{\lower.5ex\hbox{\ltsima}}

\begin{document}

\title{ The Rotation Curve of the Large Magellanic Cloud \\
and the Implications for Microlensing}

\bigskip

\author{David R.~Alves$^{1}$ }
\affil{Space Telescope Science Institute, Baltimore, MD 21218\\
        Email: {\tt  alves@stsci.edu} }

\author{Cailin A.~Nelson$^{1}$ }
\affil{Department of Physics, University of California, Berkeley, CA 94720 \\
        Email: {\tt  cailin@physics.berkeley.edu} }

\altaffiltext{1}{also affiliated with the Institute of Geophysics and Planetary
Physics, Lawrence Livermore National Laboratory, Livermore, CA 94550}

\clearpage

\begin{abstract}

The rotation of the disk of the Large Magellanic Cloud (LMC)
is derived from the radial velocities of 422 carbon stars 
(Kunkel, Irwin, \& Demers 1997, A\&AS, 122, 463).
New aspects of this analysis include the propagation of uncertainties
in the LMC proper motion with a Monte Carlo, and a self-consistent
modeling of the rotation curve and disk kinematics.
The LMC rotation curve reaches a maximum circular velocity of
72~$\pm$~7~km~s$^{-1}$ at $R = 4.0$~kpc and then declines.
The rotation curve
is well fit by a truncated, finite-thickness exponential disk model
with no dark halo, implying
a total mass of 
5.3~$\pm$~1.0~$\times$~$10^{9}$~$M_{\odot}$.   
The velocity dispersion 
in concentric radial bins from $R = 0.5$ to 5.6~kpc
decreases from
22 to 15~km~s$^{-1}$, then
increases to $\sim$20 km~s$^{-1}$ at larger radii.
Constant-thickness disk models in virial equilibrium
cannot be reconciled with the data
even if the effects of LMC or Galactic
dark halos are included.   
If the disk is virialized,
the scale height rises from 
$h$ = 0.3 to 1.6 kpc 
over the range of $R$~= 0.5 to 5.6 kpc.
Thus the LMC disk is flared.
We model the 
velocity dispersion at large
radii ($R >$ 6 kpc) as a maximal flared disk under
the influence of the Galactic dark halo, which
favors a mean density for the latter of 
$\overline{\rho} \sim 2.5 \times 10^{-4}$~$M_{\odot}$~pc$^{-3}$
at the LMC distance.  

LMC stellar kinematics play an important role in elucidating
the nature of MACHOs, a dark population inferred from LMC microlensing.
We have constructed a truncated and flared maximal
disk model for the LMC which is kinematically based.  
Our model does not
include a nonvirialized component such as tidal debris.
The instantaneous probability of microlensing
from LMC stars in our model
is $\tau < 1.0 \times 10^{-8} \cdot 
{\rm sec}^2 i$, where $i$ is the disk inclination.
Our upper limit on the self-lensing optical depth
is in good agreement with that obtained
from less sophisticated models,
and is an order of magnitude too small
to account for the MACHO microlensing signal.

\end{abstract}

\keywords{Galaxy: structure  ---
cosmology: dark matter ---
galaxies: Magellanic Clouds, structure ---
stars: carbon, kinematics}

\clearpage

\section{Introduction}

If the Galactic dark halo were composed
entirely of massive compact halo objects (MACHOs), 
the instantaneous probablity of microlensing (the ``optical
depth'') towards the Large Magellanic Cloud (LMC) would have been
$\tau \sim 5 \times 10^{-7}$ 
(Pacz\'{y}nski 1986).
The most recent observational result for MACHOs with
masses up to about 1 $M_{\odot}$ is
$\tau = 1.2^{+0.4}_{-0.3} \times 10^{-7}$,
which comes from an analysis 
of 5.7 years of LMC microlensing survey data (Alcock et al.~2000).
Thus an all-MACHO Galactic dark halo is ruled out.
However, the observed
optical depth is significantly higher than the estimate
for known Galactic and LMC stellar populations, 
$\tau \simlt 4 \times 10^{-8}$ (Alcock et al.~2000).
The excess LMC microlensing signal may be telling us
the MACHO fraction of the Galactic dark halo or revealing
gaps in our understanding of the essential structure of the LMC 
and Galaxy.

Critical discussions of the 
microlensing optical depth from known stellar
populations
have developed along two lines.  One debate has been about the
possible existence of a ``new'' stellar population which would account 
for the LMC microlensing result.   Some example suggestions
include an intervening dwarf galaxy or tidal stream
(Zhao 1998; Zaritsky \& Lin 1997, Zaritsky et al.~1999), 
and a very warped Galactic disk (Evans et al. 1998).
These hypotheses have been tested, and in some cases ruled out
(Alcock et al.~1997; Beaulieu \& Sackett 1998; Bennett 1998; 
Gould 1998, 1999; 
Gyuk, Flynn, \& Evans~1999; 
Ibata, Germaint, \& Beaulieu 1999;
Johnston 1998).
The second debate 
has been about the 
importance of LMC star-star ``self-lensing'' 
(Sahu 1994; Wu 1994).  
In particular, although the LMC is well represented by an
exponential disk model which often serves as the basis for
the self-lensing optical depth calculation, 
it may exhibit some important
detailed structures. 
Our paper is motivated by this
LMC self-lensing problem.

The line-of-sight velocity dispersion of LMC stars yields a strong
constraint on star-star self-lensing from the virialized LMC 
disk: $\tau \simlt 1 \times 10^{-8}$ (Gould 1995).
Gould's elegant limit is probably
uncertain by no more than a factor of $\sim$2.
However, several LMC
models have been devised which would
increase the self-lensing optical depth over the Gould limit.
One example of a detailed structure that might
increase the self-lensing optical depth 
is a highly-inclined and flared LMC disk (Zhao 1999).  
It has also been suggested that the oldest LMC disk stars
have a very large characteristic scale height, 
or that the LMC harbors an as yet unseen 
but massive stellar spheroid\footnote{The distinction between
a massive stellar spheroid or very thick
disk and an LMC dark halo is a matter 
of the characteristic mass to light ratios
(i.e.~M/L~$\sim$~2-4 for the former, and a very large M/L for the latter),
and the density profiles.}
(e.g.~Aubourg et al.~1999; Salati et al.~1999; Evans \& Kerins 2000).
Finally, it is possible that a nonvirialized stellar component
(i.e., a shroud of
tidal debris; Weinberg 2000) acts to increase the
self-lensing optical depth (Zhao 1998).

Kinematic studies
play a critical role in testing these various proposed LMC models.
For example, the flare of
the LMC disk may be inferred 
from its velocity dispersion at different radii,
and the existence of a
non-equilibrium stellar component lying near the
LMC might be proven with a
large kinematic survey (Zhao 1999; Graff et al.~1999; see also
Zaritsky et al.~1999; Ibata et al.~1999).
In addition, one could search directly for an LMC 
stellar population with a
velocity dispersion of $\sim$50 km~s$^{-1}$, the prediction
for a spheroid or very thick disk.  These latter components must be
identified before their importance to the self-lensing optical depth
can be assessed.
The LMC rotation curve provides
a framework for detailed kinematic studies such as these.

The decomposition of the LMC rotation curve
into disk and dark halo components
has several implications for microlensing.
First, as emphasized by Gyuk et al.~(1999), 
the total mass of the LMC
is an important constraint on
self-lensing optical depth calculations.
A high-mass model will typically predict a high
self-lensing optical depth, with an additional dependence on whether 
the mass lies mostly in the disk or in a halo.
Unfortunately,  
recent analyses of the LMC rotation curve have lead to 
mass estimates that range over a
factor of $\sim$4, which can be
attributed primarily to 
different assumptions about a dark halo
(Kim et al.~1998, Kunkel et al.~1997, Schommer et al.~1992).
Second, 
a massive LMC dark halo might significantly affect the
LMC disk kinematics (e.g.~Bahcall 1984), with possible
consequence for constraints
on self-lensing or determining the flare of the disk.
The influence of an LMC dark halo on the LMC disk kinematics
has not been investigated in detail.
We note that the interplay between
the disk and dark halo
is also of general interest for studies
of galaxy formation and evolution. 
Finally, if the LMC
has a dark halo with a 
physical makeup similar to that of
the Galactic dark halo (i.e.~composed partly of MACHOs), the 
MACHO fraction of the Galactic dark halo implied by the observed optical
depth would be lowered (Alcock et al.~2000).

Motivated by the above considerations, we present a
new analysis of the LMC rotation curve.  
In this work
we are particularly concerned with the flare of the LMC disk.  
However, other issues pertinent to microlensing are discussed as our
calculations and analyses permit.
We refer extensively to the \ion{H}{1} rotation curve recently presented
by Kim et al.~(1998) and the
impressive radial velocity dataset
for LMC carbon stars summarized by 
Kunkel, Demers, Irwin \& Albert (1997; hereafter KDIA).
A significant subset of these latter data 
are public (Kunkel,
Irwin \& Demers 1997), and archived electronically at the 
CDS\footnote{Centre de Donnes astronomique de Strasbourg; located
at http://cdsweb.u-strasbg.fr.}.  We also refer extensively to the 
analysis of the LMC space motion presented by Kroupa \& Bastian (1997).

Our paper is organized as follows.
In \S2, we present the impetus for this work, a
comparison of constant-thickness exponential
disk models (with no dark halos) to the LMC disk velocity dispersions
reported by KDIA.
In \S3, we reanalyze the LMC rotation curve and velocity
dispersion curve using archived carbon-star radial-velocity data.  
In \S4, we present a multi-mass component kinematic model for the LMC.
In \S5, we compare our carbon-star rotating disk solution to our model.
We discuss the microlensing implications
of our analyses in \S6 and
conclude in \S7.

\section{A Flared LMC Disk?}

\subsection{Theoretical Models for Disk Galaxies}

Disks in galaxies often have exponential surface brightness profiles
from which we infer exponential mass surface density profiles
of the form
\begin{equation}
\Sigma(R) = \Sigma_{0} \ e^{-R/\Lambda}
\end{equation}
by assuming constant mass-to-light ratios ($M/L$).  In equation~(1),
$\Sigma_{0}$ is the central mass surface density, $\Lambda$ is the radial
scale length, and $R$ is a cylindrical radial coordinate.  For an infinitely
thin disk described by equation (1), Freeman (1970) calculated that the 
rotation curve reaches a maximum circular velocity at $R \sim 2.2\Lambda$
\begin{equation}
V_{max} = 0.623 \  \left(\ \frac{G\  M_{disk}}{\Lambda} \ \right)^{1/2}
\end{equation}
where $M_{disk}$ is the total mass of the disk, which is related to
the central surface density by
\begin{equation}
M_{disk} \ = \ 2\pi \Sigma_{0} \Lambda^{2}
\end{equation}
For the case of a disk with a modest finite thickness, 
$V_{max}$ will be decreased
by $\sim 5\%$ (van der Kruit \& Searle 1982).   A useful model with a
finite thickness is the exponential disk (also known
as a double-exponential disk),
which has a density distribution
\begin{equation}
\rho(R,z) = \rho_{0} \ e^{-|z|/h} \ e^{-R/\Lambda}
\end{equation}
In equation~(4),
the variable $h$ is the vertical, or 
what we will sometimes refer to as the ``z'' scale height.  
If we choose the spatial density normalization
$\rho_{0} =
\Sigma_{0}/2h$, the exponential 
disk projects to the surface density of 
the infinitely thin Freeman disk, i.e.~equation~(1).
The z-velocity dispersion, $\sigma_z$, for the exponential disk 
is given by Wainscoat, Freeman and Hyland (1989),
\begin{equation}
\sigma_z(R,z) = 2h \ \left[ \pi G \ \rho_{0} \ e^{-R/\Lambda}
\left( 1 - \frac{1}{2}e^{-|z|/h} \ \right) \ \right]^{1/2}
\end{equation}
which assumes that the disk is virialized.
Equation~(5) is strictly valid for a
radially infinite disk. 
The radial dependence (i.e., the $e^{-R/\Lambda}$ term) in equation~(5) 
scales the local density normalization, 
as prescribed by others
(van der Kruit \& Searle 1982; 
van der Kruit \& Freeman 1984).

A model extensively discussed by
van der Kruit \& Searle (1981, 1981b, 1982) 
is the isothermal disk.
The vertical distribution of stars in an isothermal disk 
is described by the sech$^2$ function, as originally derived
by Spitzer (1942; see also \S4.2 of this paper).
The spatial density for this disk model is,
\begin{equation}
\rho(R,z) = \rho_{0} \ {\rm sech}^2(z/h) \ e^{-R/\Lambda}
\end{equation}
which projects to the surface density of the infinitely thin
Freeman disk if $\rho_{0} = \Sigma_{0}/2h$.  
Assuming virialization, the z-velocity dispersion
of the isothermal disk is proportional to the z-scale height
and independent of z:
\begin{equation}
\sigma_z(R) = h \ \left[ 2 \pi G \ \rho_{0} \ e^{-R/\Lambda} \ \right]^{1/2}
\end{equation}
where the radial dependence 
scales the local density normalization.  
Equation~(7) is strictly valid
for a radially infinite disk.  We note that the ``z-scale height'' 
in the isothermal disk is not strictly a scale height (and not
the same as $h$ in the double-exponential model), but it is a similar parameter.
Further comparisons of the exponential and isothermal disk models
are found in Wainscoat, Freeman \& Hyland (1989), and
van der Kruit (1988).

The systematic error in the velocity dispersion predicted by
the radially infinite isothermal disk model has been shown
to be only $\sim$10\% in the 
range $1 < R/\Lambda < 4$ using numerical studies of
truncated isothermal disk models
(van der Kruit \& Searle 1982).  We will use the isothermal
disk model in our analyses,
because these numerical calculations 
provide us with an estimate of the model accuracy, 
and will later allow us to approximately account for the
finite radial extent of the LMC disk
(see \S4.3 of this paper).  Observations
of edge-on spirals indicate that both the exponential and isothermal
models are viable representations of real disks
(de Grijs, Peletier, \& van der Kruit 1997).
The specific choice of model
does not affect the main results of this work.

It is reasonable to compare
the model z-velocity dispersion 
to the observed velocity dispersion 
at different projected radii in the LMC 
for the case of uniform anisotropy ($\sigma_Z/\sigma_R$ constant
at different radii; van der Kruit \& Freeman 1984).
Although we do not know if this
is true for the LMC, the assumption of uniform anisotropy
is supported by kinematic studies of the old Galactic disk 
(Lewis \& Freeman 1989).  Gould (1995) notes that 
if $\sigma_Z/\sigma_R <$ 1 as found elsewhere,
the line of sight velocity dispersion measured for an inclined disk
overestimates $\sigma_Z$.  Therefore, our model
z-velocity dispersion is likely an upper-limit
on the true z-velocity dispersion.

Combining equations (2), (3) and (7) yields
\begin{equation}
\sigma^2_z(R) \ \approx \ 1.288 \ \left(\frac{h}{\Lambda}\right) 
\ V^2_{max} \ e^{-R/\Lambda}
\end{equation}
which is most accurate for disks in the range $1 < R/\Lambda < 4$.  
The z-scale height has been observed to be constant 
over a large range of $R$ in numerous disk galaxies
(van der Kruit \& Freeman 1984; de Grijs, Peletier, \& van der Kruit 1997),
implying that $\sigma$
decreases with increasing $R$ in these galaxies.  Indeed, this trend 
of decreasing velocity dispersion has been observed
in over a dozen spiral/disk galaxies (Bottema 1993), including our own.

In summary, if the disk velocity dispersions are the same at different
radii, the disk is likely flared (e.g.~Zhao 1999).
However, this inference assumes that
(1)~the disk is virialized,
(2)~the radial scale length is constant,
(3)~the disk velocity dispersions exhibit uniform anisotropy,
and (4)~the galaxy has no dark halo (or that the dark halo
has a negligible affect on the
disk velocity dispersions).  Finally, one must be careful to compare
the observed velocity dispersions to the simple model described
above over a restricted range 
of radii, because real galactic disks are finite in radial
extent.  We compare to the LMC in the next section.

\subsection{Comparison of Disk Models to the LMC}

In this section, we will adopt a plausible value for
the product $h V_{max}^2$ in equation (8), 
and compare that model to LMC disk velocity dispersion data.
Following KDIA and many others,
we adopt a distance to the LMC of 50.1 kpc, and an inclination
of $i$ = 33 degrees.  For comparison,
wide-field UV polarimetric image data have yielded a precise
but model-dependent
estimate of the inclination of the LMC disk: 
$i = 36^{+2}_{-5}$ degrees
(Cole, Wood, \& Nordsieck 1999).
Westerlund (1997) summarizes
a number of other inclination measurements.

We adopt $\Lambda = 1.6$ kpc
for the radial
scale length of the LMC disk.
The surface brightness profile of the LMC disk 
is well-fit by an exponential with a scale length
$\Lambda \approx 1.6$~kpc 
(de~Vaucouleurs 1957, Bothun \& Thompson 1988). 
Moreover, the surface density profile 
of intermediate-age long period variables
(LPVs) in the LMC is also fit by an exponential with a
scale length of $\Lambda \sim 1.6$~kpc (Hughes, Wood \& Reid 1991).
This result directly associates our kinematic dataset with
an LMC population that follows the $\Lambda$ = 1.6~kpc exponential
profile, because some of the intermediate-age LPVs
are also carbon stars\footnote{The KDIA carbon stars
have a mean velocity dispersion similar to that of the
Hughes et al.~(1991) intermediate-age LPV sample.  Therefore, as
KDIA also note, these two samples of stars probably have similar ages 
and conform to the dynamics of the same inclined disk.}.
Finally, the RR Lyrae stars in the LMC
also appear to lie in a $\Lambda \approx 1.6$~kpc
exponential disk (Alcock et al.~2000b).  
Since the RR~Lyraes are presumably
much older than the carbon stars, 
the carbon stars were likely
born into the same disk.
We note that no stellar population
in the LMC has yet been shown to be inconsistent with a
$\Lambda = 1.6$~kpc exponential profile.  We assume 
that the carbon stars observed by KDIA 
properly represent this LMC disk.

The KDIA dataset consists of radial velocity measurements
for 759 carbon stars spanning a true LMC radius 
of 2 to 10 kpc.  
KDIA summarize the results of 11 different
zonal solutions for a rotating disk.  
In Figure~1, we plot the KDIA velocity dispersions 
as a function of radial distance, which 
we take as the central
value for each zonal solution given in KDIA's Table~1.   
We show data only for $R$ = 3 to 6 kpc.  This is
the range of radii where equation~(8)
will most accurately predict the velocity dispersions.   
The dashed line in Figure~1 is the mean of
the plotted, observed velocity dispersions, $\sigma =$ 13.7 km~s$^{-1}$.  
Assuming a constant value of $\sigma$
equal to the mean, we find $\tilde\chi^2$ = 0.4, consistent with the data.
The solid line plotted in Figure~1 shows the prediction
of equation~(8) assuming values of $V_{max} = 70$ km s$^{-1}$
and $h$ = 0.35 kpc at $R~=~2\times\Lambda$.
This model predicts 
$\sigma_z~\approx~13.7$~km~s$^{-1}$ at $R~=~2\times\Lambda
\approx~3.2$ kpc.  The fit for this model distribution
is $\tilde\chi^2$ = 8.7, which can be rejected
with high significance.  Models with larger scale heights intersecting
the data at larger $R$ give
similarly poor fits\footnote{These values of $\tilde\chi^2$ are not
strictly apppropriate because the zonal solutions given by KDIA
are not all independent.  However, 
subsets of the $\sigma(R)$ measurements which are
independent also give poor fits to the distribution given by equation~(7).}.
One possible interpretation of the observed
constancy of $\sigma$ with $R$ is that the z-scale height 
increases as $\sim e^{+R/2\Lambda}$, i.e.
{\it the LMC disk is flared.}

\section{The LMC Rotation Curve Revisited}

An alternate interpretation of 
Figure~1 might be a constant-thickness
disk under the dynamical influence 
of a dark halo (Bahcall 1984).  
Indeed, Kim et al.~(1998) compared
their \ion{H}{1}
rotation curve with the KDIA carbon star
rotation curve and, on this basis,
argued that an
LMC dark halo is dynamically significant 
at a radius of $\sim$4 kpc.
However, KDIA favored a different interpretation; they
attributed the rising outer portion
of their carbon star rotation curve to tidal effects, 
and not a dark halo\footnote{In any case,
the effect of the LMC dark halo (if it exists)
would be small at the radius we chose to normalize the model
$\sigma(R)$ curve in Figure~1, supporting this aspect of our
analysis in \S2.}.
Before concluding that the LMC disk is flared, it is worth
testing disk plus dark halo kinematic models
in a self-consistent manner.
With this model comparison in mind, let us consider the KDIA
analysis in greater detail.

We have several specific concerns with the KDIA analysis.
First, and perhaps most importantly,
KDIA excluded some carbon stars from their solutions
for their association with a polar ring.  This
procedure might have
artificially lowered
the disk velocity dispersions.  By retaining these stars
in our solutions, we guarantee that our results will be properly
comparable to our model
(which will not include a polar ring).
KDIA also corrected for the LMC transverse
motion by forcing the position angle of the kinematic
line of nodes to match the photometric line of nodes in each 
zonal solution.  This is a questionable procedure.  Moreover,
it is inconsistent
with the \ion{H}{1} analysis by Kim et al.~(1998).
We will correct the carbon-star velocities for the transverse motion 
of the LMC in a manner consistent with
the \ion{H}{1} rotation curve analysis by Kim et al.~(1998).
In fact, we will
make our carbon star rotation curve analysis
similar in other ways to the \ion{H}{1} rotation curve analysis 
(i.e., by adopting the
same kinematic center), in order 
to lend maximum
credibility to detailed comparisons of the two rotation curves.

\subsection{Rotation Curve Solution}

\subsubsection{The Carbon Star Radial Velocity Data}

We assemble positions 
(right ascension and declination; equinox 1950) and
galactocentric radial velocities for Magellanic
Cloud carbon stars from Kunkel, Irwin \& Demers (1997). 
We discard all Small Magellanic Cloud (SMC) carbon stars,
``inter-cloud'' carbon stars (their Table~15), and
carbon stars located near the center
of the LMC (their Table~17; see also \S3.3).  We also discard one
rogue carbon star (C0433$-$6607) for its highly discrepant
velocity.  The resulting homogeneous 
dataset contains 422 carbon stars, representing
a significant subset of the data analyzed in KDIA. 
Ongoing observational campaigns to obtain carbon star radial
velocities will likely increase the sample size
in the coming years (Suntzeff 1998).
The archival dataset we have
assembled is sufficient for the purposes of this study.

\subsubsection{Transverse Motion Radial Velocity Correction}

Feast, Thackeray \& Wesselink (1961) have discussed the
apparent rotation induced by the transverse motion of the LMC. 
Each of our 422 carbon star velocities are corrected for this projected
radial velocity gradient as follows.
We adopt the LMC space motion calculated by Kroupa \& Bastian (1997),
which is derived from an average
of the LMC proper motion they measure with {\it Hipparcos} data and the LMC proper
motion measured by Jones, Klemola, \& Lin ~(1994).  
The {\it Hipparcos} and Jones et al.~(1994) measurements 
of the LMC proper motion agree within their respective
error bars.  We note that Kim et al.~(1998) make their transverse motion
correction using the proper motion measurement by Jones et al.~(1994).

It is useful
to establish a Galactic coordinate system for the vector
analysis that follows.  Following Kroupa \& Bastian (1997),
the galactocentric coordinates used here are such that the
Galactic Center and the Sun are located at 0 and $-8.5$ kpc
along the $x$-axis, respectively.  The positive $z$-axis 
points toward the north Galactic pole, and the positive
$y$-axis points in the direction of Galactic rotation at
the position of the Sun.  In these coordinates, the position
vectors of the Sun and LMC are
\begin{equation}
R_{\odot} \ = \ \left(-8.5, \ 0.0, \ 0.0 \right) 
\end{equation}
\begin{equation}
R_{LMC} \ = \ \left(-1.0, \ -40.5, \ -26.6  \right) 
\end{equation}
in units of kpc.  The velocity vector of the LMC is
\begin{equation}
V_{LMC} \ = \ \left(+41\pm44, \ \ -200\pm31, \ \ +169\pm37 \right) 
\end{equation}
in units of km s$^{-1}$
(Kroupa \& Bastian 1997).
The galactocentric radial velocity
of the LMC is the vector dot product $V_{LMC}~^.~\hat{R}_{LMC} = 
(41, -200, -169)~^.~(-1.0, -40.5, -26.6)/48.5 =  74$ km s$^{-1}$,
where $\hat{R}_{LMC}$ is a normalized unit vector.
The vector between the Sun and LMC is $X = R_{LMC} - R_{\odot} =
(7.5, -40.5, -26.6)$, and the radial velocity component of the
LMC seen from the local standard of rest (accounting for the Sun's
peculiar velocity and Galactic rotation) at the position of the Sun
is $V_{LMC}~^.~\hat{X}_{LMC}
= 80$ km s$^{-1}$.   The vector $X_{LMC}$ points toward the 
``center'' of the LMC, which is assumed to be at 
$\ell = 280.46$, $b = -32.89$ (Kroupa \& Bastian 1997).
The unit vector connecting the Sun to any 
position in the sky in galactocentric coordinates is
\begin{equation}
\hat{X}(b,\ell) \ = \ \left( \cos~b \cos~\ell, \ \ \cos~b \sin~\ell, 
\ \ \sin~b \right)
\end{equation}
The transverse motion velocity correction that 
we seek is the difference between the
projected radial velocity toward the LMC center and the projected
radial velocity toward each carbon star.  If we define the
vector $\hat{S} = \hat{X}_{LMC} - \hat{X}(b,\ell)$, the apparent
rotation induced by the LMC space motion is $\delta V(\ell,b) = 
\hat{S}~^.~V_{LMC}$.  We make our correction by subtracting
$\delta V(\ell,b)$ from the observed carbon star
velocities.   Small differences between the Kroupa \& Bastian (1997)
model used here for the transverse motion radial velocity correction
and assumptions (i.e.,~for the LMC center and distance) in subsequent
sections have a negligible effect on our results.

\subsubsection{Zonal Solutions}

First, we convert each carbon star's right
ascension and declination into spherical coordinates in
units of radians.  We designate these 
$\alpha$ and $\delta$, respectively.
We adopt the center of the
LMC at right ascension $05^{h} \ 17^{m}.6$ and declination
$-69^{0} \ 02^{'}.0$ (equinox 2000) following Kim et al.~(1998).  
We designate the LMC center with
$\alpha_{0}$ and $\delta_{0}$
(also in units of radians and precessed to equinox 1950). 
The distribution of carbon stars is shown in Figure~2
(where North is up and East is to the right).
The SMC is located well outside
of the boundaries of the figure in a South-Westerly 
direction (toward the lower left).  We remind that
$\alpha$ and $\delta$ are spherical coordinates plotted
on a rectilinear scale.  Therefore, the distribution of carbon 
stars shows a $\sim$cos($\delta$) distortion.

Our solution proceeds as follows.  
We calculate the position angle ($\Theta$) and radial distance
from the center of the LMC ($S$ in units of radians)
for each carbon star using the formulae
\begin{equation}
S = \cos^{-1}\left[\ \sin \delta_{0} \ \sin \delta \ + \
\cos \delta_{0} \ \cos \delta \ \cos (\alpha - \alpha_{0}) \  \right]
\end{equation}
\begin{equation}
\Theta^{\prime} = \sin^{-1}\left[
\frac{\cos \delta_{0} \ \sin (\alpha - \alpha_{0})}{\sin S}\right]
\end{equation}
\begin{equation}
\Theta = 2\pi - \Theta^{\prime}_{NE}, \ 
\Theta = \pi + \Theta^{\prime}_{SE}, \  
\Theta = -\Theta^{\prime}_{NW}, \  
\Theta = \pi + \Theta^{\prime}_{SW} \
\end{equation}
Equation~(15) assigns the correct sign and phase to the position
angle depending on which quadrant each star resides in Figure~2.
We have defined $\Theta$ to increase West of North.
We deproject each radial distance on the sky to 
true LMC radius ($R$ in units of kpc) using,
\begin{equation}
R = 50.1 \ \left| \tan^{-1} \left( \ 
 S \ \cos(\Theta + \Theta_{0}) \ \left[1 \ + \ \tan^2(\Theta + \Theta_{0}) 
\sec^2 i \right]^{1/2} \ \right) \right|
\end{equation}
We take the absolute value of the 
inverse tangent in equation~(16) to guarantee 
a positive value for $R$. 
We accommodate a twisting disk model by introducing $\Theta_{0}$.
The parameter $\Theta_{0}$ allows for a changing kinematic line of nodes 
in different radial zone solutions.  We note that $\Theta_{0}$
is sensitive to the adopted LMC transverse
velocity (Meatheringham et al.~1989).  
We define our different zones by making cuts in true radius.
Each zonal solution is derived by fitting the function
$V(\Theta)$ to the distribution of galactocentric radial velocities, 
\begin{equation}
V(\Theta) \ = \pm\left( \ \frac{V_{c} \ \sin i}
{\left[ \  1 \ + \ \tan^2(\Theta + \Theta_{0}) \sec^2 i
\ \right]^{1/2} } \ + \ V_{sys} \right)
\end{equation}
We must choose the correct sign for $V(\Theta)$
at the appropriate position angle, which is the meaning
of the `$\pm$' in equation~(17).  We choose `$+$' for
$(\Theta + \Theta_{0})$ = $-\pi/2$ to $\pi/2$, and
`$-$' for $(\Theta + \Theta_{0})$ = $\pi/2$ to $3\pi/2$.
The fitted model could
be generalized to account for a warped and decentered disk by
letting $i$ and ($\alpha_{0}$, $\delta_{0}$) vary in each
zonal solution.  However, 
we fix  $i$, $\alpha_{0}$,
and $\delta_{0}$ to the values
given above in all solutions.

We derive solutions in four radial zones which span 
$R$ = 2.5 - 5, 5 - 6,
6 - 7, and 7 - 13 kpc.   These were
chosen to provide an adequate number
of stars and phase coverage in each zone.  
The inner-most and outer-most
boundaries were chosen to be inclusive 
of the entire dataset when $\Theta_{0}$ = 0.
We begin by solving for
$V_{sys}$, $\Theta_{0}$, and $V_{c}$ that yield
a minimum dispersion about $V(\Theta)$ in each of 
our zones.  We average the values of
$V_{sys}$, excluding the outermost zone (which was sometimes
poorly constrained).
In a second round of fitting, 
we fix $V_{sys}$ to the three-zone average value,
and search only for
values of $\Theta_{0}$ and $V_{c}$ that
minimize the velocity dispersion.  This procedure was
easily reproduced, and facilitated our
bootstrap and Monte Carlo error 
analyses (described below).

The mean radii in
each final zonal solution are $<R>$ = 4.0, 5.6, 6.5, and 7.8 kpc.
The minimum velocity dispersions occurred at $\Theta_{0}$ = 
$-$37, $-$22, $-$8, and 3 degrees .  We find circular velocities 
$V_{c}$ = 72, 68, 65, and 65 km s$^{-1}$ and velocity
dispersions $\sigma$ = 17.7, 14.9, 19.3 and 20.6 km s$^{-1}$,
respectively (in order of increasing radius).
The four zonal solutions are shown in Figure~3.  
We indicate the carbon stars
with small open circles.  Each carbon star is plotted
twice at $\Theta$ and ($\Theta$ + $2\pi$). 
The best fit curve is plotted as a solid line.
The mean radius, circular velocity, and velocity dispersion are
also labeled in each panel.

\subsection{Rotation Curve Error Analysis}

Our first estimate of the errors is made with a bootstrap resampling 
analysis (Barrow, Sonada, \& Bhavsar 1984).  We create 250 artificial
datasets by randomly removing 1-10\% of the carbon stars from
our full dataset, and then find new rotating disk solutions 
for each of these. 
In a bootstrap reasampling analysis, the variances 
in the mean values of interest averaged over all
of the artificial datasets are taken to represent the internal variances
of the full dataset.    We find errors in $V_{c}$ of 1.3, 1.7,
2.3, and 2.4 km s$^{-1}$ for the four radial zones
(in order of increasing radii).  We find an
error in $\sigma$ of 0.3, 0.2, 0.4, and 0.5 km s$^{-1}$, and
errors in $\Theta_{0}$ of 1.1, 1.0, 2.4, and 1.5 degrees, respectively.

Our second calculation is designed to
estimate the errors associated with our choice of
zonal boundaries and the movement of stars between 
zones due to the changing $\Theta_{0}$ parameters.
For each of the four zones, we find new solutions for 64 different
variations in the zonal boundaries.  We varied each
boundary by no more than $\pm$ 0.25 kpc, which yielded roughly
comparable numbers of stars and mean radii in each zonal solution
(except for the outer boundary of the outer-most zone which
was varied by 1.5 kpc in steps of 0.25 kpc).
The variances of $V_{c}$, $\sigma$, and
$\Theta_{0}$ in the
zonal boundary solutions are similar to those
found with the bootstrap analysis.  Moreover, we find no
systematic trends with boundary
choice, except for the outermost zone.  In the outermost
zone, $V_{c}$ decreases systematically (from $\sim$60 to 30 km s$^{-1}$)
as the inner boundary is varied from  $\sim$6.7 to 7.2 kpc.

Finally, we perform a Monte Carlo calculation to propagate
the error on the LMC space motion to the
rotation curve.  We create 500 artificial datasets, where 
the LMC space motions are drawn randomly from a sample that
reproduces the best fit values and errors in equation~(11).  We assume
Gaussian distributions for the space motion errors.  
This calculation yields errors
in $V_{c}$ of 7.1, 8.7, 10.4 and 15.8 km s$^{-1}$,
errors in $\sigma$ of 0.1, 0.2, 0.3 and 0.4 km s$^{-1}$, 
and errors in $\Theta_{0}$ of 6.6, 4.6, 7.6 and 8.4 degrees,
in the four zones (in order of increasing radius).  Similar values
were found with as few as $\sim$250 artificial datasets,
indicating that our calculation is robust.  For most of
the parameters in our rotating disk solution, the error from
the LMC space motion dominates.
Our final solutions and adopted errors
are summarized in Table~1.

We note that there
would be no stars in common between the different solutions
if all zones had the same $\Theta_{0}$, but in our final solutions,
$\sim$5\% of the stars
are located in two zones (see Table~1).  However, since we 
have accounted for the movement of small numbers of stars across
the zonal boundaries in our error analysis, we may consider
each zonal solution as independent in subsequent
analyses.

\subsection{Velocity Dispersion in the LMC Bar}

We previously discarded 22 radial velocity
measurements for carbon stars very near the center of the LMC
(Kunkel, Irwin \& Demers 1997; their Table~17) because this
sample of stars did not span an adequate range of $\Theta$
to yield an interesting zonal rotation curve solution.  
However, the velocity dispersion of these carbon stars is
of considerable interest.  This sample of
stars at $<R> = 0.5$ kpc has a velocity dispersion of
$\sigma = 22.1$ km~s$^{-1}$.  Guided by our error analyses above,
we adopt an error of 1 km~s$^{-1}$.
The average velocity dispersion of carbon stars 
in our five zones
is $\sigma$ = 18.9 km s$^{-1}$.
We defer further discussion and interpretation of the LMC rotation
curve and disk velocity dispersions until \S5, after we present our
theoretical kinematic model.

\section{A Multi-Mass Component Model for the LMC}

\subsection{Rotation Curve}

We will consider two representations of the contribution of 
the luminous mass in the LMC disk to the rotation curve.
The first is derived empirically by Kim et al.~(1998)
from the $R$-band surface brightness data of de~Vaucouleurs (1958).
For two assumptions of the disk mass-to-light ratio, we present
this stellar rotation curve in Table~2.  
The second representation of the disk rotation curve we will consider
is the finite-thickness, truncated disk model of van~der~Kruit and Searle
(1982).  We adopt their model No.~2 (truncated at $R$ = 4$\Lambda$) from
their Appendix A.1.  Table~3 summarizes
this disk model rotation curve, which is 
given in units of the radial scale length ($\Lambda$) and the
equivalent maximum circular velocity for an infinitely thin
exponential disk ($V_{max}$; see \S2.1 of this paper).
This model will also be used to calculate a correction to 
the scale heights inferred from the disk velocity dispersions
at small radii in the LMC (see \S4.3 of this paper).

When comparing either of these two disk
rotation curves to our carbon star data points, we will
add in quadrature the small additional contribution of gas in 
the LMC disk to the rotational velocity.  
Kim et al.~(1998) calculated the gas 
rotation curve using the
single-dish \ion{H}{1} data of Luks \& Rolfs (1992),
and assuming a 30\% contribution from \ion{He}{1}.
The gas rotation curve is summarized in Table~4.

Finally, we will account for a pseudo-isothermal dark halo in the traditional
manner, assuming a density profile,
\begin{equation}
\rho_{halo}(R) \ = \ \frac{ \rho_{0} }{1 \ + \ \left(R^2/a^2\right) }
\end{equation}
where the circular velocity is a function of the integrated
total mass
\begin{equation}
V^{2}_{halo}(R) \  =  \ 
4 \pi G \rho_{0} a^2 \left[
1 \ - \ \frac{a}{R} \tan^{-1} \left(\frac{R}{a}\right) \ \right]
\end{equation}
The psuedo-isothermal dark halo is parameterized by a
central density, $\rho_{0}$, and core radius, $a$.
We add the circular velocities of the 
disk ($V_{disk}$) and psuedo-isothermal halo ($V_{halo}$)
in quadrature in order
to calculate the rotation curve of our model: $V_{c}(R)$.

\subsection{Disk Velocity Dispersions}

The effect of a spherical dark halo on the velocity dispersion
of an embedded disk has been extensively discussed by Bahcall (1984),
Bahcall \& Casertano (1984) and Bottema (1993).  The 
vertical motions of stars in a disk and halo system are
governed by Poisson's equation
\begin{equation}
\frac{\partial^2\phi}{\partial z^2} \ = \ 4 \pi G 
\left( \rho_{disk} \ + \ \rho^{eff}_{halo} \right)
\end{equation}
where
\begin{equation}
\rho^{eff}_{halo} \ = \ \rho_{halo} \ -  \ 
\frac{1}{4 \pi G R} \  \frac{\partial}{\partial R} \ V^{2}_{c}(R)
\end{equation}
and the first moment of the Boltzmann equation
\begin{equation}
<\sigma^2_z> \ \frac{\partial\rho(z)}{\partial z} 
\ = \ -\frac{\partial\phi}{\partial z} \ \rho(z)
\end{equation}
The effective halo density, $\rho^{eff}_{halo}$,
includes the contribution from the radial deriviative
of the total circular velocity, $V_{c}$.
If the rotation curve is fairly flat,
we may discard the radial deriviative term in equation (22)
and equate $\rho^{eff}_{halo} = \rho_{halo}$.
This formulation of the problem (Bahcall 1984)
neglects the $<\sigma_z \sigma_R>$ cross terms in 
Poisson's equation.
With the following redefinitions (Bahcall 1984)
\begin{equation}
z_{0} = \left( \frac{<\sigma^2_z>}{2 \pi G \rho_{disk}(z=0)}
\right)^{1/2}, 
\ \ \  x = \frac{z}{z_{0}} \ \ , \ \ \
y(x)  = \frac{\rho_{disk}(x)}{\rho_{disk}(x=0)} 
\end{equation}
and
\begin{equation}
\epsilon  = \frac{\rho^{eff}_{halo}(z=0)}{\rho_{disk}(z=0)}
\end{equation}
we rewrite the equations governing the vertical motions of
the disk stars as
\begin{equation}
y \frac{d^2 y}{dx^2} \ - \ \left(\frac{dy}{dx}\right)^2 \ = \
-2y^2 \ - \ 2 \epsilon y^2
\end{equation}
with the boundary conditions
\begin{equation}
y(0) \ = \ 1  \ , \ \ \ \left(\frac{dy}{dx}\right)_{x=0} \ = 0
\end{equation}
The solution for $\epsilon$ = 0 was first published by Spitzer (1942):
\begin{equation}
y \ = \ {\rm sech}^2 x
\end{equation}
The solution to equation~(25) for $\epsilon >> 1$ is
\begin{equation}
y \ = \ e^{-\epsilon x^2}
\end{equation}
The differential equation~(25)
must be solved numerically for intermediate values of $\epsilon$.

The $\epsilon$ parameter
define in equation (24) is the
ratio of effective halo density to the disk density in the plane of
the disk.  It is particularly useful parameter because it relates
ratio of the velocity dispersion in a disk plus halo
system, $\sigma_{disk+halo}$, to the velocity dispersion
in a disk with no halo, $\sigma_{disk}$.  
Following Bottema (1993),
and suppressing the subscript $z$ for clarity,
\begin{equation}
\frac{\sigma_{disk+halo}}{\sigma_{disk}}  \ = \ \alpha^{-1} 
\end{equation}
where $\alpha$ is the surface density
\begin{equation}
\alpha = \int\limits^{\infty}\limits_{0}  y(x) dx
\end{equation}
In the limit of $\epsilon >> 1$, 
\begin{equation}
\alpha^{-1}  = \  2\left(\frac{\epsilon}{\pi} \right)^{1/2} 
\end{equation}
We have solved for $\alpha^{-1}$ as a function of $\epsilon$
numerically\footnote{Bottema (1993) also made this calculation.
We note that Bottema's Fig.~15 is incorrect.  However, the relevant
text in Bottema (1993) is correct.};
these values are listed in Table~5.  
In the limit of large $\epsilon$, these values confirm equation~(31).
The error in $\alpha$ is $\sim$30\% at $\epsilon \sim 1$ and
less than 20\% at $\epsilon > 2$, in the sense that equation~(31)
overestimates the numerically calculated value of $\alpha$.

Finally, we note that the
density profile of the dark halo may be derived from the observed
velocity dispersion in the disk at large $R$, if the disk scale height
is constant.  Equating $\rho^{eff}_{halo} = \rho_{halo}$ and
$\sigma_{obs}(R) = \sigma_{disk+halo}(R)$, and combining
equations~(6), (7), (24), (29), and (31) gives
\begin{equation} 
\sigma^2_{obs}(R) \ = \ 8 \ G \ h^2 \ \rho_{halo}(R,z=0) \ \approx \ 
8 \ G \ h^2 \ \frac{\rho_{0} \ a^2}{R^2}
\end{equation} 
where the approximate density profile of the dark halo (at large $R$)
follows from equation~(18).
Thus, if a galactic disk is embedded in a dark halo,
one expects $\sigma_{obs}(R) \propto R^{-1}$ at large $R$
with the slope proportional to $a \rho^{1/2}_{0}$.

\subsection{The Numerical z-Force Correction}

The velocity dispersion in the LMC bar will yield 
an accurate estimate of
the scale height only if we make a
numerical correction to account for the
finite extent of the LMC.  We note that
applying a pure disk model at small
radii in the LMC is reasonable 
because the LMC has no bulge, and the
surface brightness profile (even in the bar) 
is purely exponential (Bothun \& Thompson 1988).  
We appeal to the numerical integrations by 
van~der~Kruit \& Searle (1982) to estimate
a ``correction factor'' for the disk velocity dispersions
given by equation~(7).
In Table~A1 of van~der~Kruit \& Searle (1982), model No.~2,
the ratio of the actual z-force ($K_z$) to the model z-force 
is 0.55, 0.82, 0.95, 1.07, 1.06, 0.64, and 0.77 at
$R/\Lambda$ = 0, 1, 2, 3, 4, 5 and 6, respectively.
We have simply averaged over the tabulated
values of $z/z_{0}$ (various other weighted-average schemes
give similar values).  The
correction to the model velocity dispersons
is the square root of the z-force ratios
(van~der~Kruit \& Searle 1982), and is $\sim$25\%
near the LMC center.  The scale height inferred from
the velocity dispersion in the LMC bar would be underestimated
by a factor of $\sim$2 without the correction.

\subsection{Summary of Theoretical Calculations}

Here we summarize the theoretical
calculations presented thus far.
In \S2.1, we introduced the basic formulae
that describe an exponential disk with no dark halo, 
and a prediction for the 
velocity dispersion perpendicular to the plane 
of the disk.  The latter calculation 
was made by assuming a radially infinite disk.
For a disk truncated at $R/\Lambda$ = 4 or 5,
the error in this predicted velocity dispersion 
is less than 10\% at radii of a few scale lengths
(van~der~Kruit \& Searle 1982).
In \S4.1, we 
described the rotation curve of an exponential
disk embedded in a psuedo-isothermal dark halo.  
We presented a stellar
rotation curve derived from LMC surface brightness data, 
and a gas rotation curve, both
from Kim et al.~(1998).  We presented a model rotation curve
for a finite-thickness disk truncated at $R/\Lambda$ = 4, from
van~der~Kruit \& Searle (1982).
In \S4.2, we
calculated the velocity dispersion perpendicular 
to the plane of a disk that is embedded in a 
psuedo-isothermal dark halo.  At large radii, the dynamical
influence of the dark halo can dominate.  In this limit,
we showed that the disk velocity dispersion can be predicted
by an analytic formula.  We estimated the error of this approximation
using numerical integrations. 
In \S4.3,
we give a numerical correction to the 
velocity dispersion perpendicular
to the plane of the disk (the prediction from \S2.1)
that accounts for truncation of the
disk at $R/\Lambda$ = 4, 
which is adopted from van~der~Kruit and Searle (1982).
Although the LMC disk may not be truncated at precisely
four radial scale lengths, by using the same model 
throughout this work, our analysis is self-consistent.

\section{Analysis of the LMC Kinematic Structure}

\subsection{Decomposition of the Rotation Curve}

In Figure~4, we present the LMC rotation curve and a
``maximal'' disk decomposition.  
Our four zonal solutions are
indicated with bold dots and error bars.  
Data points on the \ion{H}{1} rotation curve from Kim et al.~(1998) are shown
with bold cross symbols.
For comparison, we plot the carbon star zonal solutions of KDIA
with open circles.
Also shown in Figure~4 are the gas and stellar disk
rotation curves.
We assume M/L = 2.2 for the stellar rotation curve, 
which is similar to that found for the
Galactic disk (Bahcall, Flynn \& Gould 1992).
Disk mass-to-light ratios of $\sim$2 are
favored by dynamical stability arguments (Bottema 1993).
Finally, we plot the sum
of the gas and stellar disk rotation curves as a
bold solid line.

Our four zonal solutions and the solutions of KDIA agree within 
our respective errors.
The model rotation curve 
and our data points show quite good agreement.  In fact, although
the overall scaling of the model is a free parameter (the M/L ratio
of the stellar disk), the variation with radius is reproduced
remarkably well.   The disagreement between the model and \ion{H}{1} data points
would appear less severe if error bars similar to those on the carbon
star data points were appropriate.
If the lower envelope of the
carbon star error bars were more representative of
the true LMC rotation curve, then one could adopt a slightly
smaller M/L ratio and the agreement with the \ion{H}{1} data points
would improve.   However, since the
\ion{H}{1} data points appear to deviate from a smooth curve,
most notably the dip at $R\sim$ 3 kpc, it is possible that
all of the \ion{H}{1} data points lying below the model curve 
represent a significant failure of the model.
We speculate that 
a bar in the LMC may be responsible, although we have
not attempted any modeling of this effect.

In Figure~5, we show the same carbon star and \ion{H}{1} 
data points as in Figure~4.  We now replace the
stellar rotation curve derived from surface brightness data with
the finite-thickness, truncated model disk rotation curve. 
The gas rotation curve, and the sum of the gas and
model disk rotation curves are also plotted.  
The model curve assumes a scale length
of 1.6 kpc and an
infinitely thin disk equivalent maximum circular 
velocity\footnote{We clarify that the maximum
circular velocity in our rotation curve is 72 km s$^{-1}$.  A finite-thickness
disk has a maximum circular velocity approximately 5\% smaller
than an equivalent infinitely thin disk (see Table~3).
In this decomposition, we adopt the finite-thickness disk model, but account
for the contribution from gas (approximately 5\% near maximum), and quote here
the maximum circular velocity for an equivalent infinitely thin disk.}
$V_{max}$ = 71 km~s$^{-1}$.  
The model disk rotation curve is clearly consistent
with the data, and fits the run of \ion{H}{1} and carbon star data points
even better than the (semi-empirical) stellar
rotation curve shown in Figure~4.  The small differences between
the model disk rotation curve, the stellar disk rotation curve, 
and the \ion{H}{1} data points as illustrated in Figures~4 and 5
are not critical to our conclusions.

We estimate the mass of the LMC disk using 
$V_{max}$ = 71 km s$^{-1}$ and the formulae in \S2.1,
which yields
$M_{disk} = 4.8\pm1.0$~$\times$~$10^{9} M_{\odot}$.  The
error is estimated by adopting the uncertainty of
the maximum observed circular velocity in our carbon star solutions
($\pm$7 km sec$^{-1}$; see Table~1) for the uncertainty of
$V_{max}$.  This maximal disk model has
a surface density normalization of
$\Sigma_{0} = 298\pm59$ $M_{\odot}$ pc$^{-2}$. 
Kim et al.~(1998) estimate the total mass of gas in the LMC 
to be 0.5~$\times$~$10^{9} M_{\odot}$.  Thus, the total mass of 
the LMC is $5.3\pm1.0$~$\times$~$10^{9} M_{\odot}$.

In Figure~6, we present a ``minimal'' disk 
decomposition of the LMC rotation curve.  
We plot the same carbon star and \ion{H}{1} data points
as in Figures~4 and 5.  We plot the stellar rotation curve
assuming M/L = 1, the gas rotation curve, and the sum of the stellar
and gas rotation curves.  We also
plot the contribution of a pseudo-isothermal dark halo,
and the sum of the disk and halo rotation curves.
In this decomposition, we calculate
$M_{disk} = 1.1\pm1.0$~$\times$~$10^{9} M_{\odot}$ and
$\Sigma_{0} = 68$ $M_{\odot}$~pc$^{-2}$.  The shallow and
declining run of carbon star data points favors
a small core radius for the LMC dark halo.  The dark halo shown
has $a$~=~1~kpc and $\rho_{0}$~=~0.10~$M_{\odot}$ pc$^{-3}$.
By assuming M/L = 1 in the disk, this
decomposition has a maximal halo, and yields
a total LMC mass 
of $\sim$5~$\times$~$10^{9} M_{\odot}$
(corresponding to an upper limit on the LMC global mass-to-light
ratio of $\sim$4).

\subsection{Decomposition of the Disk Velocity Dispersions}

In Figure~7, we plot our LMC disk velocity dispersions 
as a function of true radius with bold dots and error bars.
We plot the results of KDIA 
with open circles (see also Fig.~1).  
The prediction of equation~(8) for our maximal
disk model is shown as a dotted line and labeled.
We plot this model prediction again, but now corrected
for the truncation of the LMC disk (the ``$K_Z$ force correction'')
as a bold solid line.   We have assumed a constant scale height
of $h = 0.5$ kpc, which normalizes the latter curve to 
intersect the carbon star data point in the LMC bar.  
A constant-thickness disk model cannot be reconciled with the data
by any choice of $h$.

For our minimal disk decomposition, we must account for the 
effect of the LMC dark halo on
the disk velocity dispersions. 
By combining equations (6), (18), and (24), and equating
$\rho_{halo} = \rho^{eff}_{halo}$, we may estimate
the $\epsilon$ parameter,
\begin{equation}
\epsilon(R) = \frac{\rho_{0,halo} \left(1 \ + \ \frac{R^2}{a^2}\right)^{-1}}
{\rho_{0,disk} \ e^{-R/\Lambda} }
\end{equation}
Assuming a constant scale height 
of $h = 0.4$ kpc, the spatial density 
normalization of the minimal
disk is $\rho_{0,disk}$ = 0.068 $M_{\odot}$ pc$^{-3}$.
Adopting the parameters of the maximal LMC dark halo from above, 
we find $\epsilon \approx$ 1.1, 1.0, 1.2, 1.9 and 3.4 at integer
steps of $R/\Lambda$ = 1 to 5.  For these values of $\epsilon$,
the error associated with equation~(31) is too large, and thus we
appeal directly to our numerical integrations in Table~5.

For this decomposition, we
begin with the velocity dispersion according
to equation~(8), and then make the $K_{Z}$
force correction. 
We calculate $\epsilon(R)$ using equation~(33), and use spline interpolations
of the numerical data in Table~5 to calculate $\alpha$.   
The model disk velocity dispersions are corrected according to equation~(29) 
by a factor of $\alpha^{-1}$ (equating $\sigma_{obs} = \sigma_{disk+halo}$).
These various curves are plotted in Figure~8.
Our choice of $h$ = 0.4 kpc normalizes the dark halo-corrected
curve to intersect the data point in the LMC bar.
A constant-thickness
disk in the presence of a maximal dark halo cannot be reconciled
with the observed velocity dispersions by any choice of $h$.  
For comparison, 
a fit to the velocity dispersion data with the minimal
disk and an arbitrary pseudo-isothermal 
dark halo implies a maximum circular velocity in the rotation curve
of $\sim$500 km s$^{-1}$, which is about 7 times higher 
than observed.

\subsubsection{Scale Heights of the Carbon Stars}

Our maximal disk model yields $h$ = 0.25, 0.93, and
1.62 kpc at $R$ = 0.5, 4.0, and 5.6 kpc, respectively.  For comparison,
the minimal disk yields $h$ = 0.55, 1.62, and 2.61 kpc
at the same radii.  At larger radii ($R >$ 6 kpc), where the
disk velocity dispersions begin to rise,
the implied scale heights for the maximal
disk model are $h$ = 5.1 and 28.4 kpc 
(at $R$ = 6.5 and 8.2 kpc, respectively).  The 
scale height inferred from the last data point 
is much larger than the
tidal radius of the LMC (Weinberg 2000), and
obviously wrong.  We will return to
the interpretation of
these high velocity dispersions at large radii in a moment.
A linear regression
on the three data points with $R < 6$ kpc yields:
\begin{equation}
\sigma(R) = -1.39 \ (\pm0.10) \times R \ + \ 22.89 \ (\pm0.42)
\end{equation}
where $R$ is in units of kpc, and $\sigma(R)$ is in units of
km s$^{-1}$.   We adopt this
regression to represent the flare of the LMC disk.
The change of scale height is
represented by a function of the form
\begin{equation}
h(R) \ = \ \kappa \ e^{+R/\beta\Lambda}  
\end{equation}
where $h$ and $R$ are in units of kpc.  
We find $\beta = 1.4$, and $\kappa = 0.14$ kpc.
Equation~(35) predicts the scale height to
within 0.1 kpc of the values estimated above.

Assuming a constant mass-to-light ratio,
our flared disk model must project to a
surface density that varies as $e^{-R/\Lambda}$
in order to reproduce the observed surface brightness 
profile of the LMC
(de~Vaucouleurs 1957, Bothun \& Thompson 1988).
Therefore,
the spatial density of the flared disk, $\rho_{f}(R)$, must change 
with an effective radial scale length that compensates for
the variation of scale height with $R$.  This effective radial scale length
is easily calculated by considering
\begin{equation}
\Sigma(R) \ = \ \Sigma_{0} \ e^{-R/\Lambda} \ = \ 
2 \ \rho_{f}(R) \ h(R) \ = \ 
2 \  \rho_{0 f} e^{-R/\gamma\Lambda} \ \kappa e^{+R/\beta\Lambda}
\end{equation}
where we require that $(1/\gamma - 1/\beta) = 1$.  Substituting 
$\beta$ = 1.4 yields $\gamma$ = 0.583 (note that $\gamma$
and $\beta$ are dimensionless scale factors).  It also follows that
$\rho_{0 f}$ = $\Sigma_{0}/(2\kappa)$ = 
1.064 $M_{\odot}$ pc$^{-3}$.  In summary, our
flared disk model has the following spatial density,
\begin{equation}
\rho_{f}(z,R) \ = \ \rho_{0 f} \ e^{-R/\gamma\Lambda} \ 
{\rm sech}^2\left(\frac{z}{h(R)}\right)
\end{equation}
where $h(R)$ is given in equation~(35).

\subsubsection{Dynamical Influence of the Galactic Dark Halo}

The dynamical influence of the Galactic dark halo
is calculated following the discussion in
\S4.2 of this paper (see also Aubourg et al.~1999).  
We designate the mean density of the 
Galactic dark halo at the distance of the LMC
as $\overline{\rho}$.  
If we substitute $\overline{\rho}$ for $\rho_{eff}$
in equation~(20), then $\epsilon$ from equation~(24) yields $\alpha$
from Table~5, and thus the correction to the disk velocity
dispersions through equation~(29).
We estimate $\overline{\rho}$
following Griest (1991; see his Eqn.~[3]).  
We assume that the Galactic dark
halo has a core radius of 3 kpc, and a solar neighborhood spatial
density normalization of 0.0079 $M_{\odot}$ pc$^{-3}$. 
We adopt the distance from
the Sun to the Galactic center of 8.5 kpc, and the distance from
the Sun to the LMC of 50.1 kpc, which yields
$\overline{\rho}$~=~0.00025~$M_{\odot}$~pc$^{-3}$.
We estimate $\epsilon$ = 0.001, 0.007, 0.040, 0.224 and 1.246 
at integer steps of $R/\Lambda$ = 1 to 5.

In Figure~9, we plot the same
data points as in Figures~7 and 8.
We indicate the regression of equation~(34) as a solid line, and this
regression corrected for the effect of the Galactic dark halo as
a bold solid line.  Although the fit is
not perfect, the agreement with
the last data point is quite good.  
Thus the influence
of the Galactic dark halo on our flared disk model
is of the correct magnitude
to account for the outer-disk velocity dispersions. 
In addition, the radius where the 
Galactic dark halo begins to have a significant dynamical effect 
(i.e., where the velocity dispersions begin to rise in the disk) 
is also reasonably reproduced.    A more detailed modeling of
the outer-disk velocity dispersions is beyond the scope of this
paper.

We find that $\epsilon$ is quite small at all radii of interest
for the constant-thickness maximal disk model (accounting for the
Galactic dark halo but no LMC dark halo), and
for the constant-thickness minimal disk model 
(accounting for both the Galactic
and LMC dark halos).  
In summary, constant-thickness exponential
disk models cannot be reconciled with the observed run of LMC disk
velocity dispersions.

As discussed above, 
we adopted the mean density of the Galactic dark halo
at a distance of 50 kpc of
$\overline{\rho}$~=~0.00025~$M_{\odot}$~pc$^{-3}$, or
$\log \rho_{50} \sim -3.6$ (Griest 1991).  
It is worth considering the range of allowed Galactic
dark halo profiles. 
Assuming flat rotation curves at large radii,
many authors have made Galaxy models with
pseudo-isothermal density profiles 
(e.g.~ Caldwell \& Ostriker 1981;
Bahcall, Schmidt, \& Soneira 1982).
Typical core radii range from 2 to 8 kpc.
The average value 
of the density at 50 kpc predicted by these models
is $\log \rho_{50} = -3.6$, with a
standard deviation of 0.2 dex.
As a final check, we compare the N-body cold dark matter simulation
of the Galactic dark halo presented by Dubinski (1994).
The initial density profile for this dark halo model was
similar to 
an ellipsoidal Hernquist potential, 
which is like the universal dark halo profile found in dark matter-dominated
galaxies
(Kravstov et al.~1998).  Dubinski's (1994) initial
density profile was then compressed by the Galactic potential 
(e.g.~Blumenthal et al.~1986) and in its final form
predicts $\log \rho_{50} \sim -3.6$.
The density profile of the Galactic dark
halo assumed for MACHOs (Alcock et al.~2000) is also pseudo-isothermal;
thus our analysis of the LMC disk kinematics supports this dark halo 
profile over the range of interest
($\sim 8$ to 50 kpc).

\subsubsection{Tidal Debris?}

It has been suggested that the LMC
is embedded in a
``shroud'' of
tidal debris (Weinberg 2000).  
Such debris is unlikely
to contribute significantly to the LMC self-lensing optical
depth because too little mass is involved, and it is likely
to be located close
to the disk (Weinberg 2000; see \S6.2).
Nonetheless, it is worth considering the possibility
that our sample of true disk carbon stars is contaminated by
tidal debris, which might reconcile a
constant-thickness disk model with the disk kinematic data.
In this case, tidal debris would also affect 
our interpretation
of the outer-disk velocity dispersions.
The notion of tidal debris surrounding the LMC
is somewhat similar to the suggestion by KDIA
that the LMC harbors a polar ring, also of tidal origin.
(We note that KDIA analysed a larger sample of carbon stars,
in which they claim the kinematic signature 
of a polar ring is evident.)  We have thus far presented an interpretation
of the radial velocity data for 422 carbon stars in the LMC
that does not require a polar ring, or tidal debris.
The nature of tidal debris (in a polar ring or otherwise)
is investigated as follows.

If we assume that a constant-thickness maximal disk model represents 
the true LMC disk,
the ``excess'' velocity dispersion in each
zonal solution may be calculated by subtracting the model contribution
from the observed dispersions in quadrature,
which yields $\sigma_{ex}$ =
15.0, 13.7, 18.8 and 20.5 km s$^{-1}$ at $R$ = 4.0, 5.6, 6.4
and 8.2 kpc, respectively.  
Next, if we assume that the contaminating debris has a uniform velocity
dispersion of $\sim$50 km~s$^{-1}$, the 
ratio of contaminating stars to total
stars in each zonal solution would be of order
$\sim$10-20\%.  (This ratio is
inversely proportional to the velocity dispersion assumed for the
tidal debris.)   In each of our zones, we estimate the
number of $\sim$50 km~s$^{-1}$ 
tidal debris stars would be 12, 17, 10, and 8
(in order of increasing radius), if the LMC disk were
of constant thickness.

We calculate that our four disk zones subtend relative
sky areas of 0.40/0.16/0.16/1.00, in order of increasing zone radius.
A regression of the relative numbers of tidal debris stars with
the relative zone areas shows an anti-correlation, significant
at the $\sim$1$\sigma$ level.  
Assuming that the true disk and tidal debris carbon stars are
similarly affected by incompleteness in the data, 
this is not what we naively expect
from the tidal debris scenario described by Weinberg (2000).
The two outermost zones are illustrative.  They have area
ratios that differ by a factor of $\sim$6, yet the estimated numbers of
contaminating tidal debris stars are 10 and 8.
Therefore, the contaminating debris must
have a non-uniform radial distribution 
of surface density, velocity dispersion, or both.  
It is also possible, but
even more contrived, that a spatially-dependent incompleteness in the carbon 
star dataset has conspired with a non-uniform spatial distribution of
tidal debris to yield the observed velocity dispersions.
In light of this analysis,
we prefer an interpretation of the LMC rotation curve and disk
velocity dispersions that invokes a negligible contamination from
tidal debris (i.e., our maximal flared disk model).

The unambiguous detection of nonvirialized LMC stars
and the accurate characterization of their 
spatial distribution
are clearly desirable.   Observations of such a population/structure
would be relevant to our interpretation of the LMC rotation
curve and disk kinematics, with possible additional implications for 
LMC microlensing (e.g.~Zhao 1999; Graff et al.~1999; Weinberg 2000).

\section{Microlensing Implications}

\subsection{Self-Lensing of the Flared LMC Disk}

Here we calculate the LMC self-lensing optical depth of our 
maximal flared disk model
by directly integrating the spatial density of stars given
in equation~(37).
We begin with an integral of the form
\begin{equation}
\tau(R_s,\phi_s,z_s) = \frac{4\pi G}{c^2} \ \int\limits^{D_s}\limits_0 \rho(y) \
\left(1 - \frac{y}{D_s}\right) \ y \ dy \approx
\frac{4\pi G}{c^2} \ \int\limits^{D_s}\limits_0 \rho(y) \ y \ dy
\end{equation}
which yields the optical depth to any source star in the LMC
at position $(R_s,\phi_s,z_s)$, a cylindrical coordinate system
in the plane of the disk.
The position angle $\phi$ is defined to be zero
at the near side if the inclined disk.  
The integration variable is the line of sight distance 
between the source and lens.  
We use the subscripts ``$s$'' and ``$l$'' to denote 
the coordinates of the sources and lenses, respectively.
The approximation in equation~(38) is that $y$ is
always much smaller than the distance to the source 
($D_s \approx$ 50.1 kpc),
which is reasonable for LMC self-lensing. 
Note that $y$ is simply related to $D_l$, the distance from the 
observer to the lens, and $D_s$, the distance from the observer 
to the source
\begin{equation}
y = D_l - D_s = \frac{z_l - z_s}{\cos~i}
\end{equation}
where $i$ is the inclination angle of the disk
(measured from the plane of the sky).
For completeness, we provide the following geometric identity
relating the source and lens positions to $y$:
\begin{equation}
R_l^2 \ = \ R_s^2 \ + \  y^2 \sin^2i \ + \ 2  R_s y  \cos~\phi_{s}
\end{equation}
It is necessary to integrate over the line of sight a second time,
in order to calculate a density-weighted average 
of $\tau(R_s,\phi_s,z_s)$ over the distribution of source stars
\begin{equation} 
\overline{\tau}(R,\phi) \ = \ 
\frac{ \int\limits^{\infty}\limits_{0} 
\rho(R_s,z_s) \ \tau(R_s,\phi_s,z_s) \ dD_s }{
\int\limits^{\infty}\limits_{0} \rho(R_s,z_s) \ dD_s }
\end{equation} 
which may be properly compared to the observed 
optical depth.
The integration over $D_s$ is made in practice by transforming to
a variable in our LMC cylindrical coordinate system (i.e.,
$dD_s = dz_s / \cos~i$).
We truncate the LMC disk at a conservative radius of 10 kpc.
Our results are insensitive to the choice of truncation radius
at the few percent level.
We calculate $\overline{\tau}(R,\phi)$ for multiple lines of sight,
designated by ($R,\phi$) in the plane of the disk.
Our calculation is the case of ``pure'' self-lensing (Gould 1995), 
which yields an upper limit.
The effects of dust obscuration and nonlensing
mass in the plane of the disk will lower 
the real value of the optical depth for a fixed total mass of
the LMC.  

We have calculated the self-lensing optical depths in each
of the 30 survey fields
in the 5.7-year LMC microlensing analysis by
Alcock et al.~(2000; see also Gyuk et al.~1999).
In the central-most fields, 
we find $\overline{\tau} \approx \ 1.7 \times 10^{-8}$,
while in the fields lying at the largest true radii, we 
find $\overline{\tau} \approx \ 1.2 \times 10^{-8}$.
The field-averaged optical depth for LMC disk self-lensing is:
$\overline{\tau}_{30} \ = \ 1.4 \times 10^{-8}$.

The LMC inclination dominates the
uncertainty in $\overline{\tau}_{30}$.
First, we note that the inclination is probably not
a serious concern for
the rotation curve and disk kinematics
because the shape of the $V(\Theta)$ function
(Eqn.~[15]) is fairly insensitive to $i$ over
the range of plausible values.  However,
Gould~(1995) and Gyuk et al.~(1999) show that,
to first order, $\tau$ is proportional
to the square of the mass-weighted velocity dispersion and
an inclination factor of sec$^2 i$.
Therefore, we simply rescale our calculation of
$\overline{\tau}_{30}$ for different 
inclinations,
\begin{equation}
\overline{\tau}_{30} \ < \ 1.0 \times 10^{-8} \cdot 
\sec^2 i  \\
\end{equation}
which should be quite accurate
for the LMC inclinations typically found.
As noted in \S2.2, Cole et al.~(1999)
find $i = 36^{+2}_{-5}$.  Thus their $\sim$2$\sigma$ upper limit,
$i < 38$, yields
$\overline{\tau}_{30} < 1.6 \times 10^{-8}$.  For comparison,
Bothun \& Thompson (1988) find
$i = 45$, which corresponds to
$\overline{\tau}_{30} < 2.0 \times 10^{-8}$.  The contribution
to the uncertainty in $\overline{\tau}_{30}$ from 
the velocity dispersion 
follows from Gould's (1995) formula and our
equation~(34); it 
is of order $\sim (0.42/22.89)^{2}$, or a
negligible 0.03\%.   Finally, the $\sim$10\% uncertainty
in $V_{C}$ yields a $\sim$20\% uncertainty on the
total LMC mass, but this corresponds to a small uncertainty
in the factor of $1.0 \times 10^{-8}$ in equation~(42).

The self-lensing optical depth on the near and far sides of the
minor axis at the distances spanned by the 30 fields
in Alcock et al.~(2000)
varies by $\sim$5\%
due to the inclination of the disk (e.g.~Gould 1994).
Thus the flaring of the LMC disk has only a minor effect
on the spatial distribution of the self-lensing optical depth, which
is a potential diagnostic of the lens population (e.g.~Alcock et al.~2000).
The estimate of a $\sim$5\% minor axis asymmetry 
in the optical depth would increase for larger
adopted values of the inclination.

\subsection{Discussion of LMC Self-Lensing}

Gyuk et al.~(1999) recently reviewed different self-lensing
calculations in the literature.  To first order,
self-lensing from the virialized
LMC disk is proportional to the following
combinations of derivable quantities
(Gyuk et al.~1999),
\begin{equation}
\tau \ \ \propto \ \ \left(\frac{ M_{disk} \ h }{\Lambda^2}\right) \sec^2 i 
\ \ \propto \ \ \sigma^2 \sec^2 i
\end{equation}
Our model accounts for ``second order'' effects, such as the finite
radial extent and flare of the LMC disk.
We have shown that uncertainties in the LMC proper motion contribute
a $\sim$1\% uncertainty to $\sigma$ and a $\sim$20\% uncertainty to
$M_{disk}$ (via $V_{C}$),
assuming a truncated and flared maximal disk model.  Assuming this
model, our numerical
calculation of the self-lensing optical depth (Eqn.~[42]) 
is appropriate.

The maximal disk model is consistent
with LMC dynamics.  
However, if a model of the LMC with a
dark halo is preferred, the halo will probably contribute to
LMC self-lensing in addition to the disk.  In this case, our
equation~(42) is not appropriate (see Gyuk et al.~1999).  We have
adopted a minimal LMC disk, and provided an example maximal
dark halo which is consistent with the rotation curve.
Measurements of the LMC disk mass to light ratio
would better constrain the minimal LMC disk model
(our model is extreme).
For the more complicated case of an LMC with a dark halo, constraints 
on self-lensing are weaker.  This is due partly to the uncertainty
that the LMC proper motion contributes to the rotation curve,
which constrains the total LMC mass.

The LMC self-lensing optical depth may depend systematically on the
age of the stars whose kinematics are studied.  Historically, it 
has been difficult to prove that old LMC stellar populations have
different velocity dispersions.
For example,
the CH stars in the LMC 
were originally thought to trace the elusive LMC halo population,
by analogy with CH stars in the Galaxy (Hartwick \& Cowley 1988,
Cowley \& Hartwick 1991).
However, the CH stars in the LMC are now believed to represent a much
younger population (Suntzeff et al.~1993).  
We note that our average
carbon star velocity dispersion (18.9 km s$^{-1}$)
is similar to that found for the CH stars (20 km s$^{-1}$).

According to Aubourg et al.~(1999) and Salati et al.~(1999),
the LMC could harbor an old population with a large velocity dispersion.
These authors
invoke Wielen's (1977) age-velocity dispersion calibration
from the solar neighborhood. 
We note that Weilen's 
calibration was derived for
stars younger than $\sim$3 Gyr, and extrapolation
beyond this age may not be appropriate.  Moreover,
Weinberg (2000) predicts that the 
velocity dispersions in the LMC disk,
under the influence of the Galactic tidal field, will remain constant
(or decrease) over time.

Extant observational studies 
have not yielded a clear picture of the variation
of velocity dispersion with age in the LMC.  For example,
the old globular clusters in the LMC, with ages of $\sim$12 Gyrs
(Olsen et al.~1998),
exhibit a velocity dispersion of $\sim$21-24 km s$^{-1}$
(Schommer et al.~1992; see also Freeman, Illingworth, \& Oemler 1982).
Thus, the LMC carbon stars and ancient clusters support the hypothesis
of a constant disk velocity dispersion 
(for stellar populations older than $\sim$3 Gyr), in agreement with
Weinberg's (2000) prediction.
However, the velocity dispersion of old LPVs in the LMC,
with ages of $\sim$8 Gyr (Hughes, Wood, \& Reid 1991;
Olszweski, Sunzteff, \& Mateo 1996),
is $\sim$35 km s$^{-1}$ (Hughes et al.~1991).
Further observational studies are needed, 
particularly given the provocative old LPV result.

Last, we suggest that a shroud of
tidal debris or a polar ring, if they exist,
would probably make a negligible contribution to the 
LMC self-lensing optical depth 
because the total amount of mass involved is small.  
Weinberg (2000) reached a similar conclusion.

\section{Conclusion}

The rotation of the disk of the LMC has been derived from
the radial velocities of 422 carbon stars (Kunkel, Irwin, \& Demers 1997).
We have 
propogated the uncertainty in the LMC space motion to the
LMC rotation curve
with a Monte Carlo calculation.  The associated disk velocity 
dispersions found in each rotating disk zonal solution are less
sensitive to systematic uncertainties from the LMC space motion than the
circular velocities obtained.
We note that our carbon star rotation curve is derived in a manner
consistent with the \ion{H}{1} rotation curve analysis by Kim et al.~(1998);
thus our respective results are properly comparable.

We have fit our carbon star rotation curve and the \ion{H}{1} rotation
curve with a truncated, maximal LMC disk model, yielding a total LMC mass of
$5.3\pm1.0$~$\times$~$10^{9} M_{\odot}$.
We also conclude that the disk of the LMC is flared.  Extrapolating the
flare inferred at small radii (where tidal perturbations 
are small) to the outer-disk, and accounting for
the influence of the Galactic dark halo, we are able to approximately
reproduce the observed disk kinematics.
This model favors an isothermal density profile for the
Galactic dark halo out to a distance of 50 kpc.
Our truncated and flared
maximal disk model yields a limit on the spatially-averaged
LMC self-lensing optical depth
of $\overline{\tau}_{30} \ < \ 1.0 \times 10^{-8} \cdot \sec^2 i$.
For plausible values of the LMC inclination, this low self-lensing
rate compared to the measured microlensing rate allows for the existence
of a dark lens population in the Galactic halo
(Alcock et al.~2000).

Finally, we caution that we have not included 
in our analysis (1) the dynamical effects 
of an LMC bar, (2) 
large-scale non-circular motions, (3) non-uniform anisotropy
of the disk velocity dispersions, or (4) arbitrary spatial distributions
of tidal debris (i.e., a polar ring or stars out of virial equilibrium).
Despite these shortcomings, our truncated and flared maximal
disk model successfully
accounts for the general dynamical
characteristics of the LMC,
lending inferences from the
model high weight.

\clearpage

\section{Acknowledgments}

D.R.A. acknowledges Howard E.~Bond for
support of work performed at the Space Telescope Science Institute, 
and Kem H.~Cook for support of work performed at the
Lawrence Livermore National Laboratory.
NASA Research Grant NAG5-6821 ``UV, Visible, and Gravitational
Astrophysics Research and Analysis'' and DOE Contract
W7405-ENG-48 are recognized.  
C.A.N. acknowledges support from National Physical
Science Consortium Graduate Student Fellowship.
We thank Ken Freeman, David Bennett, Tim Axelrod,
Howard Bond, Kailash Sahu, Geza Gyuk, Greg Bothun, and
the anonymous referee for their helpful discussions and comments.

\clearpage

\begin{figure}
\plotone{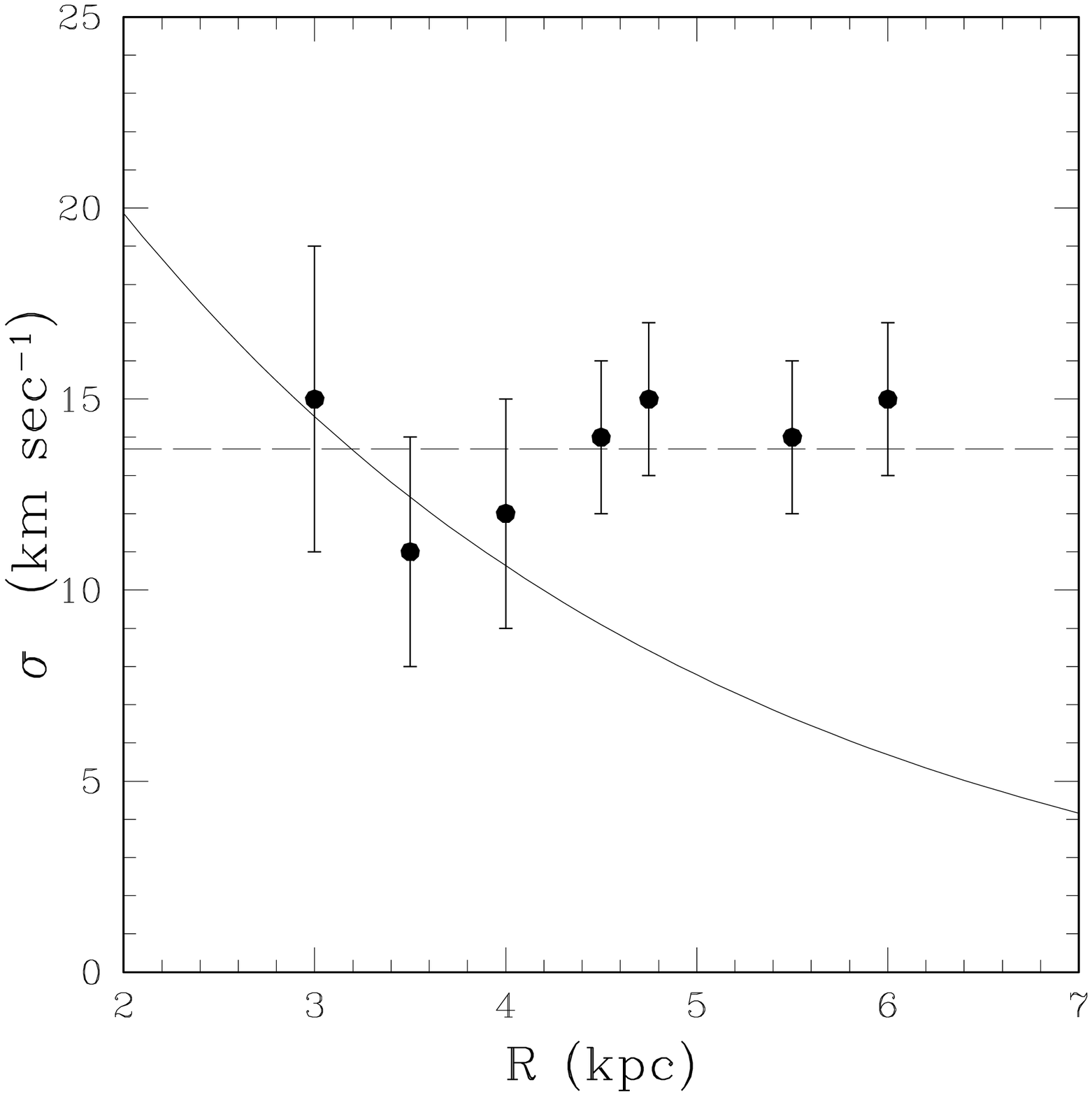}
\caption{Velocity dispersion of carbon stars in the LMC as a function
of radius.  Data taken from Kunkel, Demers, Irwin and Albert (1997; KDIA).  
Solid line shows the prediction for an exponential disk model.  
Dashed line shows the prediction for a flared disk model.}
\end{figure}

\begin{figure}
\plotone{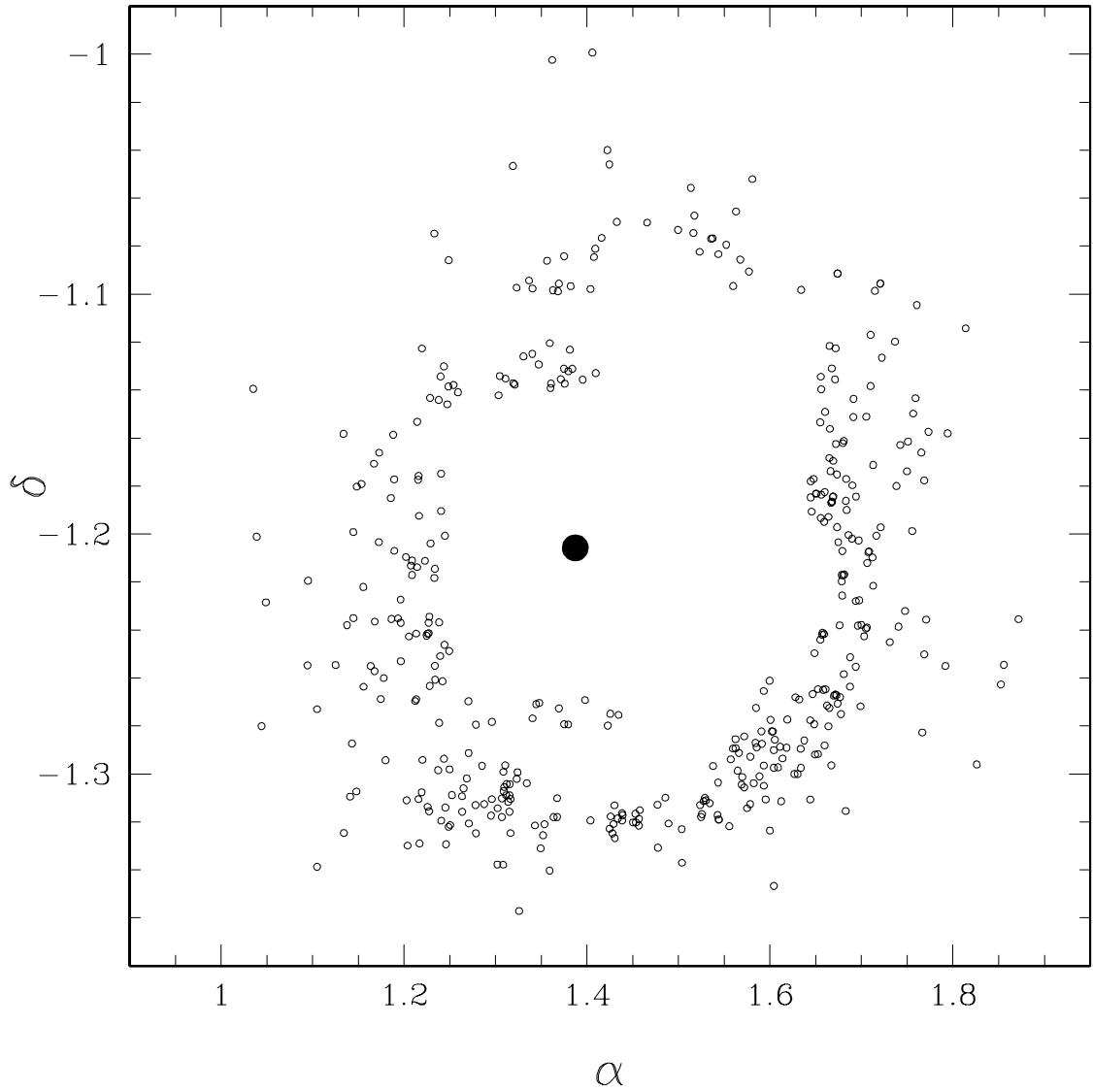}
\caption{Distribution of LMC carbon stars with archival
radial velocities (open circles) in coordinates $\alpha$ and $\delta$
(1950 right ascension \& declination in units of radians).
The bold dot indicates the kinematic center of the LMC.}
\end{figure}

\clearpage
\begin{figure}
\plotone{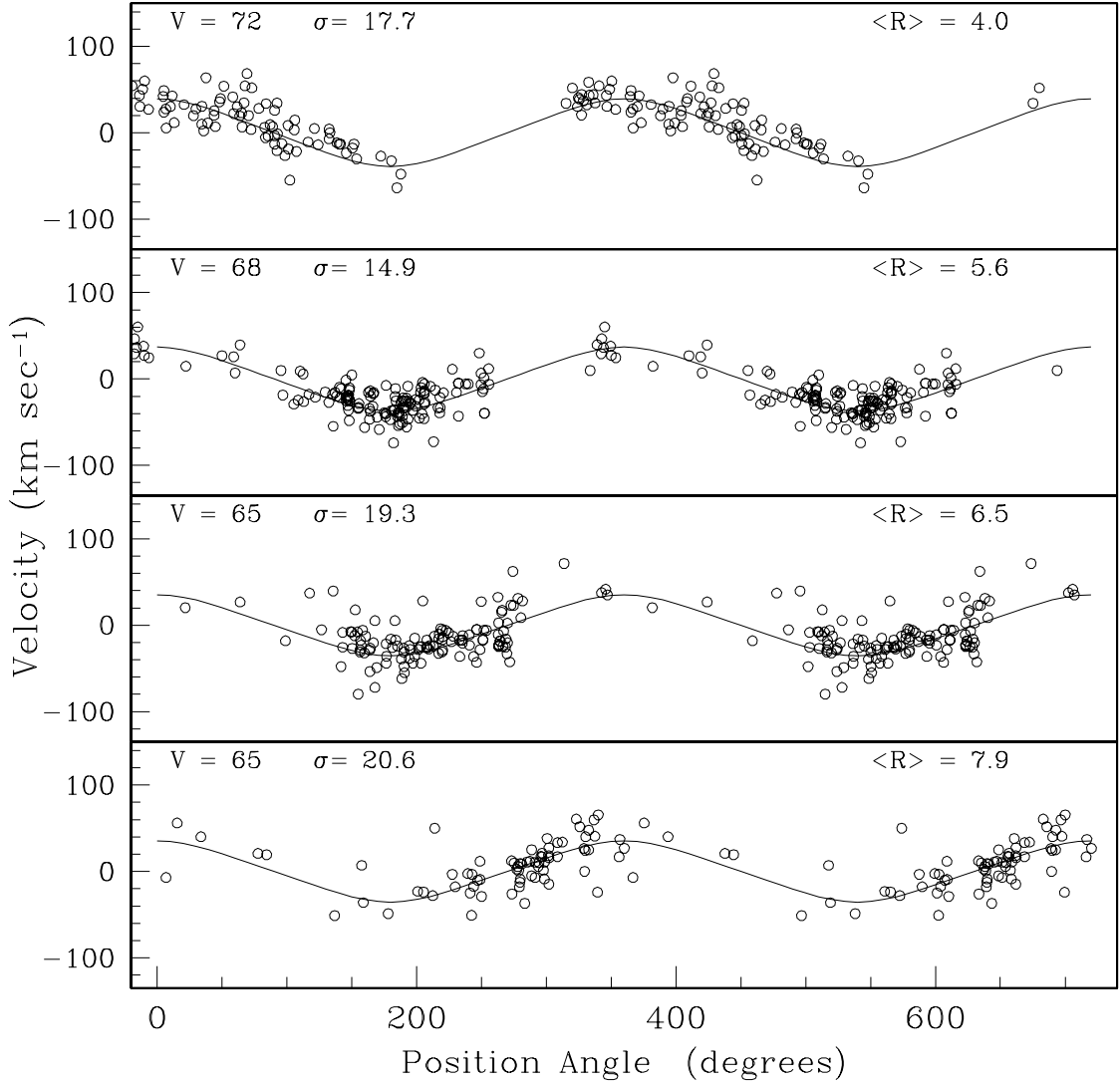}
\caption{LMC rotation curve solutions derived from carbon
star radial velocity data (open circles).  Each of the four panels
corresponds to a different radial zone.  Each carbon star is
plotted twice, at $\theta$  and $\theta + 2\pi$ for clarity.
The best fit rotating disk model is shown as a solid line through
the data.  The amplitude corresponds to the circular velocity.
The circular velocity and velocity dispersion (km s$^{-1}$), 
and mean radius
of the carbon stars (kpc) are also labeled.}
\end{figure}

\clearpage
\begin{figure}
\plotone{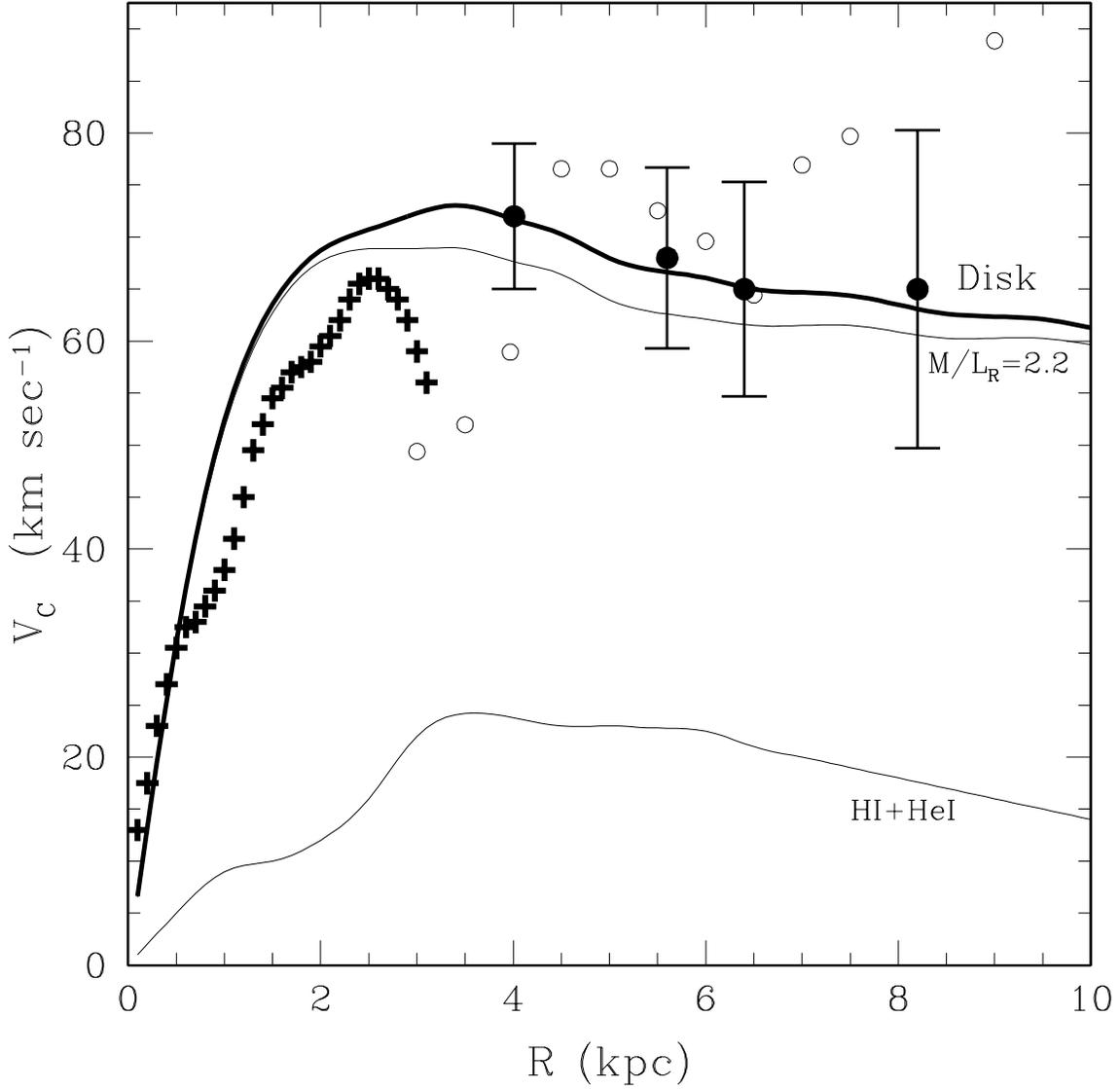}
\caption{
The LMC rotation curve: circular velocity in km s$^{-1}$ versus
true radius in kpc.  The \ion{H}{1} data of Kim et al.~(1999) are plotted with crosses.
The carbon star results of KDIA are shown as small open circles.  Our four
zonal solutions are shown as filled circles, with standard errorbars.  We plot
the gas (``\ion{H}{1}$+$\ion{He}{1}''), stellar (``M/L=2.2''), and sum of gas and stellar 
rotation curves (``Disk'').
}
\end{figure}

\clearpage
\begin{figure}
\plotone{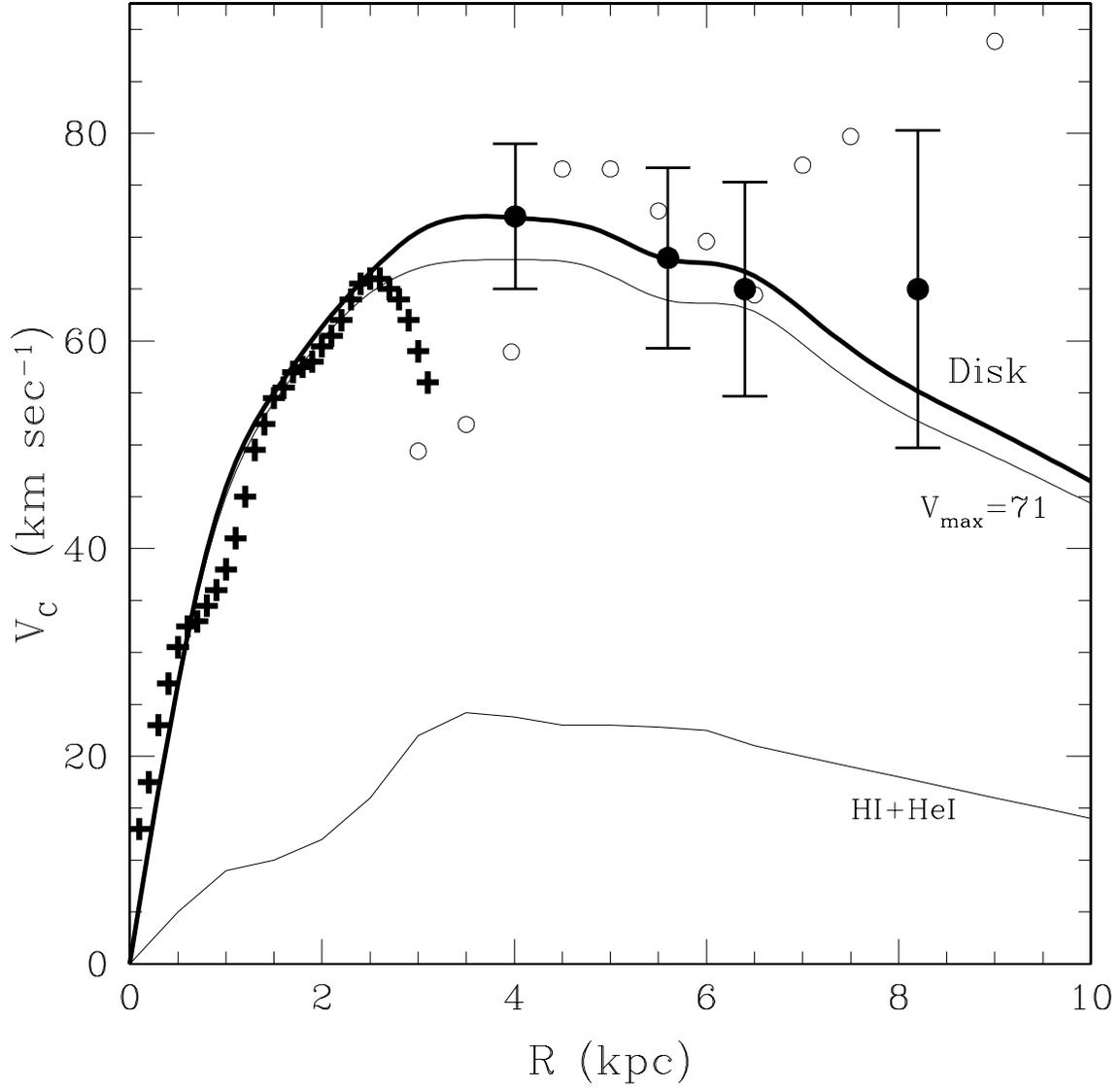}
\caption{Same as Figure~4, expect that the finite-thickness, 
truncated model disk rotation curve (``$V_{MAX}$=71'') has been substituted
for the stellar rotation curve.} 
\end{figure}

\clearpage
\begin{figure}
\plotone{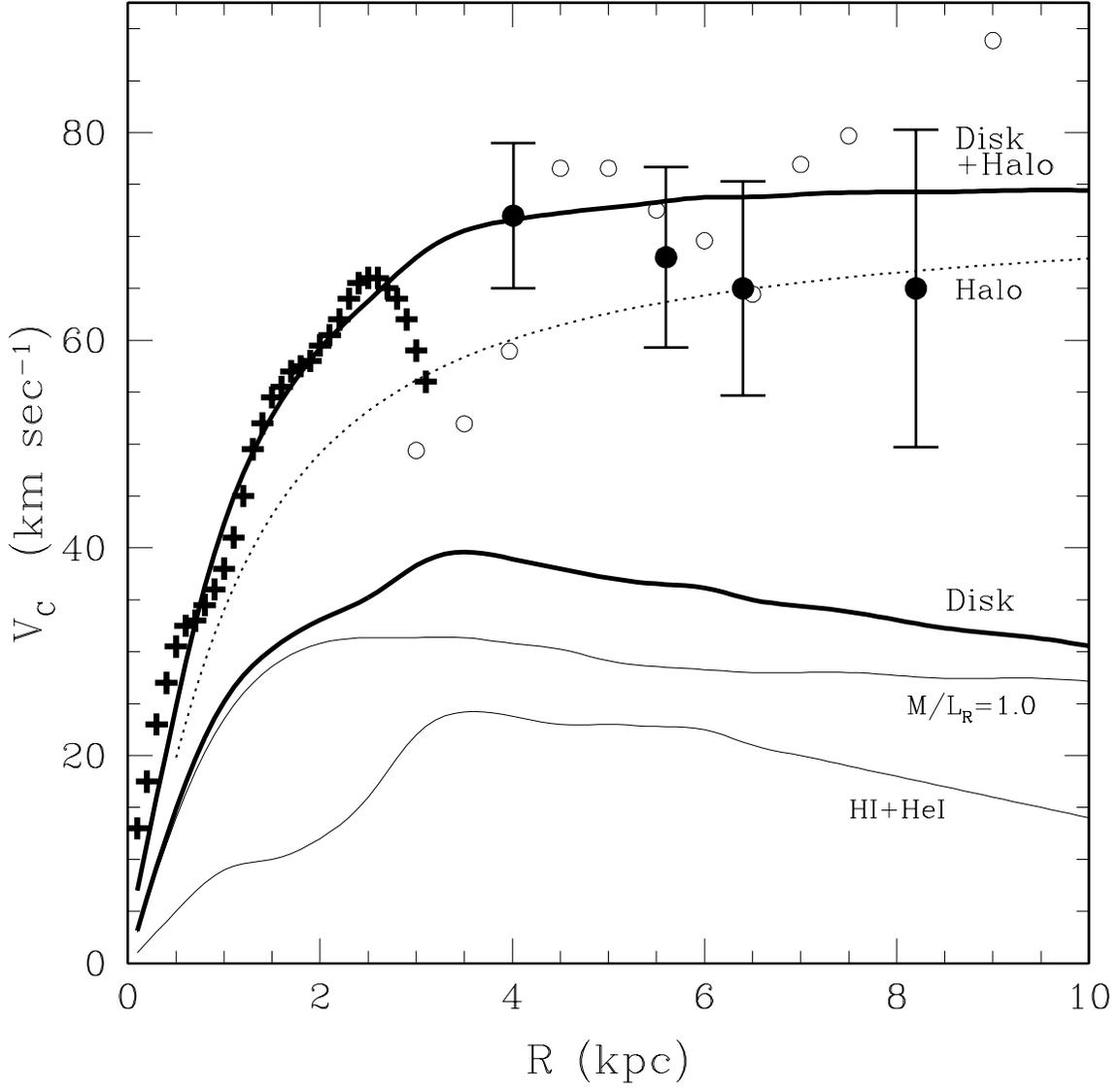}
\caption{Same as Figures~4, except that the stellar rotation curve
is now scaled with M/L=1.0, and we plot the contribution from a
psuedo-isothermal dark halo (``Halo'') and the sum of the disk
and halo curves (``Disk$+$Halo'').}
\end{figure}

\clearpage
\begin{figure}
\plotone{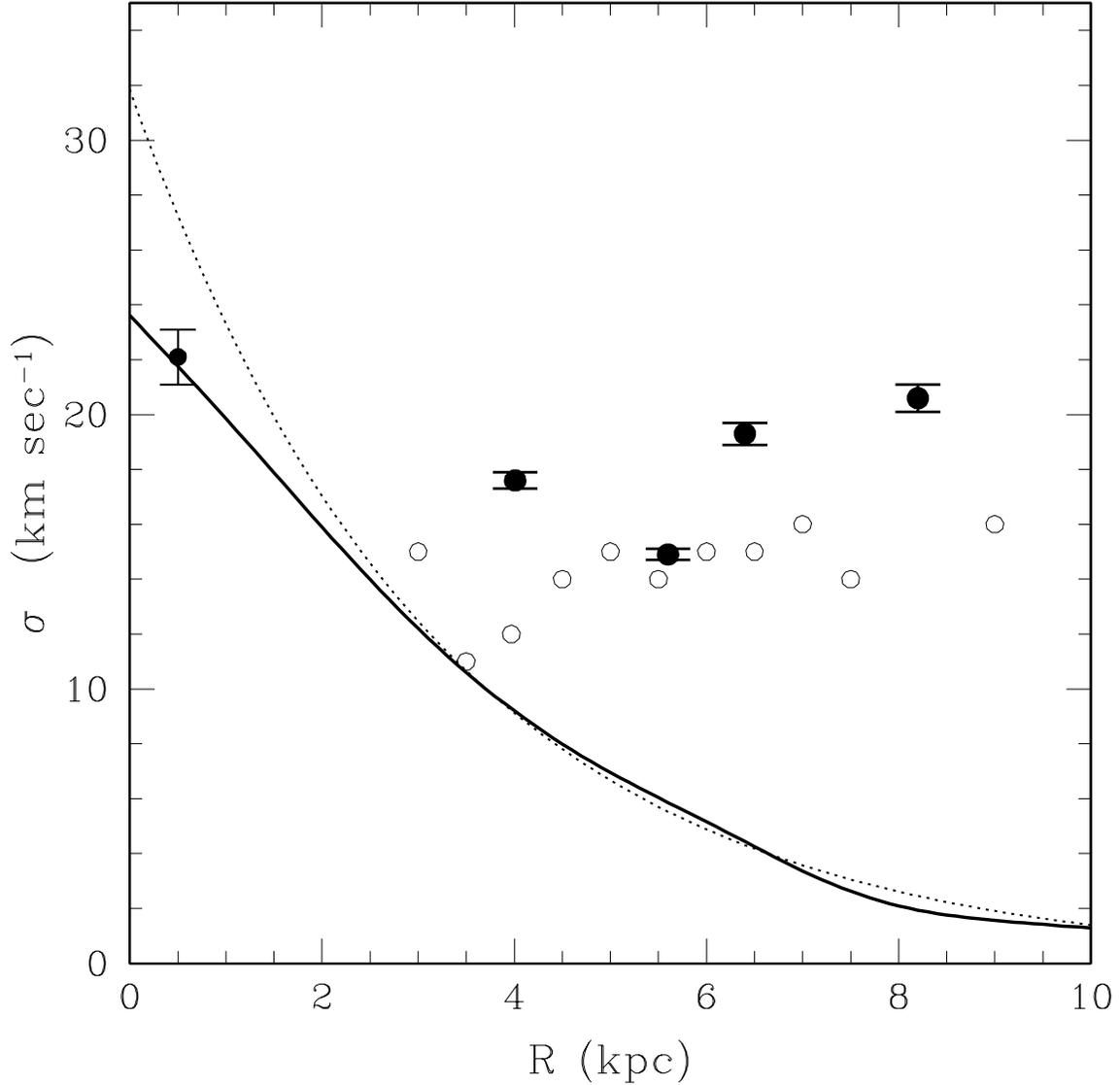}
\caption{We plot the run of LMC disk velocity dispersion with
radius (km s$^{-1}$ versus kpc).  Our points are bold dots
with errorbars.  The solutions of KDIA are shown as open circles
(errorbars omitted for clarity).  We show the predicted curve for 
our maximal LMC disk model (see also Fig.~4) as a dotted line, and this
curve properly corrected for the finite extent of the disk as a
bold solid line.  This curve has been normalized ($h = 0.5$ kpc)
to intersect the point at $R = 0.5$ kpc.}
\end{figure}

\clearpage
\begin{figure}
\plotone{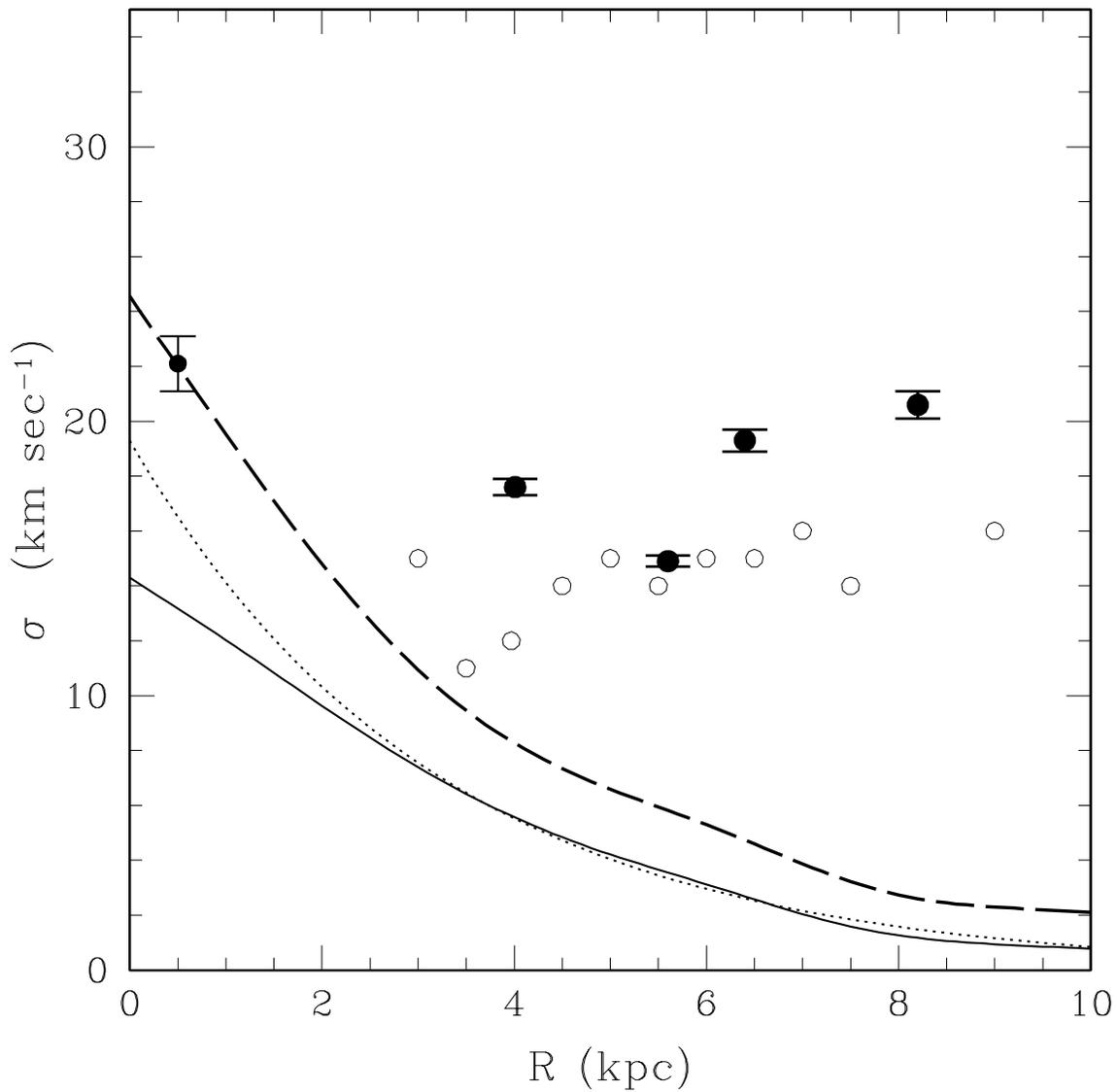}
\caption{Same as Figure~7, except that we show the prediction for
our minimal LMC disk model (dotted line), this model but corrected 
for the finite extent of the LMC (solid line), and finally this disk 
model but with the effect of a psuedo-isothermal dark halo also included
(bold dashed line).}
\end{figure}

\clearpage
\begin{figure}
\plotone{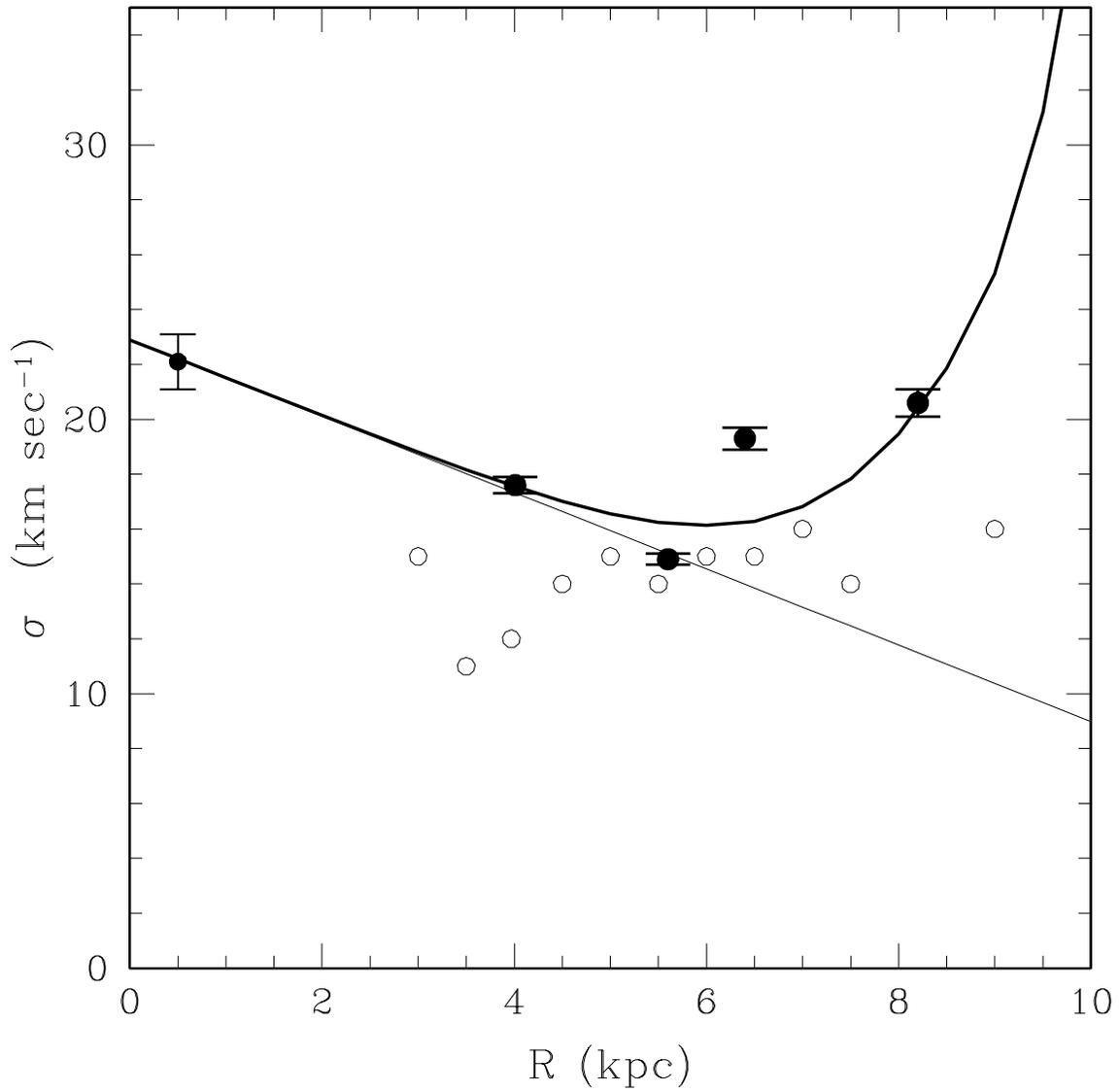}
\caption{Same as Figures~7 and 8, except that we show the prediction for
our maximal flared LMC disk model (solid line), and this model corrected 
for the effect of the Galactic dark halo (bold solid line).}
\end{figure}

\clearpage

\begin{deluxetable}{lccccclccccc}
\tablewidth11cm
\tablecaption{Carbon Star Rotating Disk Solutions}
\tablehead{
\colhead{Zone} &
\colhead{N} &
\colhead{$<R>$} &
\colhead{$\theta_{0}$ } &
\colhead{$V$} &
\colhead{$\sigma_{V}$ } & \\
\colhead{\footnotesize (kpc)} &
\colhead{} &
\colhead{\footnotesize (kpc)} &
\colhead{\footnotesize (deg.)} &
\colhead{\footnotesize (km s$^{-1}$)} &
\colhead{\footnotesize (km s$^{-1}$)} 
}
\startdata
 2.5 - 5.0 & 91 & 4.0 & $-37$ & 72 $\pm$ 7.1 & 17.6 $\pm$ 0.3  \nl
 5.0 - 6.0 & 139 & 5.6 & $-22$    & 68 $\pm$ 8.7 & 14.9 $\pm$ 0.2  \nl
 6.0 - 7.0 & 122  & 6.4 & $-8$   & 65 $\pm$ 10.4 & 19.3 $\pm$ 0.4  \nl
 7.0 - 13.0 & 73 & 8.2 & 3  & 65 $\pm$ 15.8 & 20.6 $\pm$ 0.5  \nl
\enddata
\end{deluxetable}

\begin{deluxetable}{ccc|ccc}
\tablewidth11cm
\tablecaption{Stellar Disk Rotation Curve\tablenotemark{A} }
\tablehead{
\colhead{R} &
\colhead{$V_{disk}$ \ } & 
\colhead{$V_{disk}$ \ } & 
\colhead{R} &
\colhead{$V_{disk}$ \ } & 
\colhead{$V_{disk}$ \ } \\
\colhead{\footnotesize } &
\colhead{\footnotesize $M/L_{R}$=1.0} &
\colhead{\footnotesize $M/L_{R}$=2.2} &
\colhead{\footnotesize } &
\colhead{\footnotesize $M/L_{R}$=1.0} &
\colhead{\footnotesize $M/L_{R}$=2.2} \\
\colhead{\footnotesize (kpc)} &
\colhead{\footnotesize (km s$^{-1}$)} &
\colhead{\footnotesize (km s$^{-1}$)} &
\colhead{\footnotesize (kpc)} &
\colhead{\footnotesize (km s$^{-1}$)} &
\colhead{\footnotesize (km s$^{-1}$)} 
}
\startdata
0.5 & 14.0 & 30.8 & 5.5 & 28.6 & 62.7 \nl
1.0 & 23.5 & 51.7 & 6.0 & 28.3 & 62.1 \nl
1.5 & 28.6 & 62.7 & 6.5 & 28.0 & 61.5 \nl
2.0 & 30.8 & 67.7 & 7.0 & 28.0 & 61.5 \nl
2.5 & 31.4 & 68.9 & 7.5 & 28.0 & 61.5 \nl
3.0 & 31.4 & 68.9 & 8.0 & 27.7 & 60.9 \nl
3.5 & 31.4 & 68.9 & 8.5 & 27.4 & 60.3 \nl
4.0 & 30.8 & 67.7 & 9.0 & 27.4 & 60.3 \nl
4.5 & 30.2 & 66.4 & 9.5 & 27.4 & 60.3 \nl
5.0 & 29.1 & 64.0 & 10.0 & 27.2 & 59.7 \nl
\enddata
\tablenotetext{A}{Adopted from Kim et al.~(1998).
These velocities are derived from the R-band surface brightness data
of de Vaucouleurs (1958) and assume a constant mass-to-light ratios.}
\end{deluxetable}

\begin{deluxetable}{cc|cc}
\tablewidth11cm
\tablecaption{Model Disk Rotation Curve\tablenotemark{A} }
\tablehead{
\colhead{R/$\Lambda$} &
\colhead{$V_{disk}$/$V_{max}$\tablenotemark{B}} & 
\colhead{R/$\Lambda$} &
\colhead{$V_{disk}$/$V_{max}$\tablenotemark{B}}
}
\startdata
0.5 & 0.550  & 3.5 & 0.900 \nl
1.0 & 0.780  & 4.0 & 0.890 \nl
1.5 & 0.900  & 4.5 & 0.820 \nl
2.0 & 0.950  & 5.0 & 0.750 \nl
2.5 & 0.955 & 5.5 & 0.700 \nl
3.0 & 0.945 & 6.0 & 0.650 \nl
\enddata
\tablenotetext{A}{Adopted from van der Kruit and Searle~(1982);
see their Fig.~A.1, model no.~2 (disk is truncated at 4$\Lambda$).}
\tablenotetext{B}{Circular velocities are in units of the
the maximum circular velocity of an infinitely thin exponential 
disk (see \S2.1 of this paper).}
\end{deluxetable}

\begin{deluxetable}{cc|cc}
\tablewidth11cm
\tablecaption{Gas Contribution to Rotation Curve\tablenotemark{A} }
\tablehead{
\colhead{R} &
\colhead{$V_{disk}$ \ } & 
\colhead{R} &
\colhead{$V_{disk}$}  \\
\colhead{\footnotesize (kpc)} &
\colhead{\footnotesize (km s$^{-1}$)} &
\colhead{\footnotesize (kpc)} &
\colhead{\footnotesize (km s$^{-1}$)} 
}
\startdata
0.5 & 5.0 & 5.5 & 22.8  \nl
1.0 & 9.0 & 6.0 & 22.5  \nl
1.5 & 10.0 & 6.5 & 21.0  \nl
2.0 & 12.0 & 7.0 & 20.0  \nl
2.5 & 16.0 & 7.5 & 19.0  \nl
3.0 & 22.0 & 8.0 & 18.5  \nl
3.5 & 24.2 & 8.5 & 17.0  \nl
4.0 & 23.8 & 9.0 & 16.0  \nl
4.5 & 23.0 & 9.5 & 15.0 \nl
5.0 & 23.0 & 10.0 & 14.5 \nl
\enddata
\tablenotetext{A}{Adopted from Kim et al.~(1998).
These velocities are derived from the \ion{H}{1} single-dish data of
Luks \& Rolfs (1992) and assume a 30\% contribution from \ion{He}{1}.}
\end{deluxetable}

\begin{deluxetable}{ll|ll}
\tablewidth8cm
\tablecaption{Epsilon and Alpha Integrations}
\tablehead{
\colhead{$\epsilon$} &
\colhead{$\alpha$} & 
\colhead{$\epsilon$} &
\colhead{$\alpha$} 
}
\startdata
  0.000 &  1.000 \  & 6.000  &  0.339 \nl
  0.250 &  0.865 \  & 7.000  &  0.317 \nl
  0.500 &  0.776 \ & 8.000  &  0.298 \nl
  1.000 &  0.658 \ & 9.000  &  0.283 \nl
  1.500 &  0.582 \ & 10.000 &  0.269 \nl
  2.000 &  0.528 \ & 12.000 &  0.248 \nl
  2.500 &  0.486 \ & 15.000 &  0.223 \nl
  3.000 &  0.453 \ & 20.000 &  0.194 \nl
  4.000 &  0.404 \ & 30.000 &  0.160 \nl
  5.000 &  0.367 \ & 50.000 &  0.124 \nl
\enddata
\end{deluxetable}

\end{document}